\documentclass[12pt,onecolumn,draftclsnofoot,letterpaper]{IEEEtran}
\usepackage[T1]{fontenc}% optional T1 font encoding
\ifCLASSINFOpdf
\else
\fi
\usepackage{amsmath}
\usepackage{amsbsy}
\usepackage{amssymb}
\usepackage{amsthm}
\usepackage{graphics}
\usepackage{epsfig}
\usepackage{graphicx}
\usepackage{psfrag}
\usepackage{color}
\usepackage{cases}
\usepackage{bm}
\newtheorem{theorem}{Theorem}
\newtheorem{lemma}{Lemma}
\newtheorem{example}{Example}

\newtheorem{definition}{Definition}
\newtheorem{corollary}{Corollary}

\newcommand{\ie}{{\em i.e.}}

\ifCLASSOPTIONcompsoc
\usepackage[caption=false,font=normalsize,labelfont=sf,textfont=sf]{subfig}
\else
\usepackage[subrefformat=parens,labelformat=parens,caption=false,font=footnotesize]{subfig}
\fi

\usepackage{url}
\IEEEoverridecommandlockouts
\begin{document}
\title{The Spatial Outage Capacity\\ of Wireless Networks}
\author{Sanket~S.~Kalamkar,~\IEEEmembership{Member,~IEEE,}
       and~Martin~Haenggi,~\IEEEmembership{Fellow,~IEEE}
\thanks{S. S. Kalamkar and M. Haenggi are with the Department of Electrical Engineering, University of Notre Dame, Notre Dame, IN, 46556 USA. (e-mail: $\lbrace$skalamka, mhaenggi$\rbrace$@nd.edu).}
\thanks{This work is supported by the US National Science Foundation (grant CCF 1525904).}
\thanks{Part of this work was presented at the 2017 IEEE International Conference on Communications (ICC'17)~\cite{sanket_icc17}.}}
\maketitle

\begin{abstract}
We address a fundamental question in wireless networks that, surprisingly, has not been studied before: what is the maximum density of concurrently active links that satisfy a certain outage constraint? We call this quantity the {\em spatial outage capacity} (SOC), give a rigorous definition, and analyze it for Poisson bipolar networks with ALOHA. Specifically, we provide exact analytical and approximate expressions for the density of links satisfying an outage constraint and give simple upper and lower bounds on the SOC. In the high-reliability regime where the target outage probability is close to zero, we obtain an exact closed-form expression of the SOC, which reveals the interesting and perhaps counter-intuitive result that all transmitters need to be always active to achieve the SOC, \ie, the transmit probability needs to be set to $1$ to achieve the SOC.
\end{abstract}

\begin{IEEEkeywords}
Interference, outage probability, Poisson point process, spatial outage capacity, stochastic geometry, wireless networks.
\end{IEEEkeywords}

\IEEEpeerreviewmaketitle

\section{Introduction}
 \subsection{Motivation}
In a wireless network, the outage probability of a link is a key performance metric that indicates the quality-of-service. To ensure a certain reliability, it is desirable to impose a limit on the outage probability, which depends on path loss, fading, and interferer locations. For example, in an interference-limited network, the outage probability of a link is the probability that the signal-to-interference ratio (SIR) at the receiver of that link is below a certain threshold. The interference originates from concurrently active transmitters as governed by a medium access control (MAC) scheme. Clearly, if more transmitters are active, then the interference at a receiver is higher, which increases the outage probability. Hence, given the outage constraint, a natural and a fundamental question, which has surprisingly remained unanswered, is ``What is the maximum density of concurrently active links that meet the outage constraint?'' To rigorously formulate this question, we introduce a quantity termed the {\em spatial outage capacity} (SOC). The SOC has applications in a wide range of wireless networks, including cellular, ad hoc, device-to-device (D2D), machine-to-machine (M2M), and vehicular networks. In this paper we focus on the Poisson bipolar model, which is applicable to infrastructureless networks such as ad hoc, D2D, and M2M networks.\vspace*{-1mm}

\subsection{Definition and Connection to SIR Meta Distribution}
Modeling the random node locations as a point process, formally, the SOC is defined as follows. 

\begin{definition}[\textbf{Spatial outage capacity}] For a stationary and ergodic point process model where $\lambda$ is the density of potential transmitters, $p$ is the fraction of links that are active at a time, and $\eta(\theta, \epsilon)$ is the fraction of links in each realization of the point process that have an SIR greater than $\theta$ with probability at least $1-\epsilon$, the SOC is 
\begin{align}
S(\theta, \epsilon) &\triangleq \sup\limits_{\lambda, p}~\lambda p \eta(\theta, \epsilon),
\label{eq:SOC}
\end{align}
where $\theta \in \mathbb{R}^{+}$, $\epsilon\in(0,1)$, and the supremum is taken over $\lambda > 0$ and $p \in (0,1]$.
\label{def:soc}
\end{definition}
The SOC formulation applies to all MAC schemes where the fraction of active links in each time slot is $p$ and each link is active for a fraction $p$ of the time. This includes MAC schemes where the events that nodes are transmitting are dependent on each other, such as carrier-sense multiple access (CSMA). In Def.~\ref{def:soc} $\epsilon$ represents an outage constraint. Thus the SOC yields the maximum density of links that satisfy an outage constraint. Alternatively, the SOC is the maximum density of concurrently active links that have a success probability (reliability) greater than $1-\epsilon$. Hence the SOC represents the maximum density of reliable links, where $\epsilon$ denotes a reliability threshold. We call the pair of $\lambda$ and $p$ that achieves the SOC as the \textit{SOC point}. 

We denote the density of concurrently active links that have an outage probability less than $\epsilon$ (alternatively, a reliability of $1-\epsilon$ or higher) as
\begin{equation}
\lambda_{\epsilon} \triangleq \lambda p \eta(\theta,\epsilon),
\label{eq:ss1}
\end{equation}
which results in $S(\theta, \epsilon) = \sup\limits_{\lambda, p}~\lambda_{\epsilon}$. Due to the ergodicity of the point process $\Phi$, we can express $\lambda_{\epsilon}$ as the limit
\begin{equation*}
\lambda_{\epsilon} = \lim_{r \to \infty} \frac{1}{\pi r^2} \sum_{\underset{\|y\| < r}{y \in \Phi}}\boldsymbol{1}\left(\mathbb{P}(\mathsf{SIR_{\tilde{y}}} > \theta \mid \Phi) > 1- \epsilon\right),
\end{equation*}
where $\tilde{y}$ is the receiver paired with transmitter $y$ and $\boldsymbol{1}(\cdot)$ is the indicator function. From this formulation, it is apparent that the outage constraint results in a static dependent thinning of $\Phi$ to a point process of density $\lambda_{\epsilon}$.

The probability $\eta(\theta, \epsilon)$ in \eqref{eq:SOC}, termed \textit{meta distribution} of the SIR in~\cite{martin_meta_2016}, is the complementary cumulative distribution function (ccdf) of the conditional link success probability which is given as
\begin{equation*}
P_{\rm{s}}(\theta) \triangleq \mathbb{P}(\mathsf{SIR} > \theta \mid \Phi),
\end{equation*}
where the conditional probability is calculated by averaging over the fading and the medium access scheme (if random) of the interferers, and the SIR is calculated at the receiver of the link under consideration. Accordingly, the meta distribution is given as
\begin{equation}
\eta(\theta, \epsilon) \triangleq  \mathbb{P}^{!\rm{t}}(P_{\rm{s}}(\theta) > 1-\epsilon),
\label{eq:meta_dist}
\end{equation}
where $\mathbb{P}^{!\rm{t}}(\cdot)$ denotes the reduced Palm probability, given that an active transmitter is present at the prescribed location, and the SIR is calculated at its associated receiver. Under the expectation over the point process, it is the typical receiver. The meta distribution is the distribution of the conditional link success probability, which is obtained by taking an expectation over the point process. In other words, the meta distribution is the probability that the success probability of the transmission over the typical link is at least $1-\epsilon$. As a result, as is standard in stochastic geometry, the calculation of the SOC is done at the typical user and involves averaging over the point process. Due to the ergodicity of the point process, $\eta(\theta, \epsilon)$ corresponds to the fraction of reliable links in each realization of the point process. Hence we can calculate the SOC using the meta distribution framework.\footnote{Note that the meta distribution provides the tool to analyze the SOC, but the meta distribution is not needed to define the SOC.}

\begin{figure}
\centering
\includegraphics[scale=0.58]{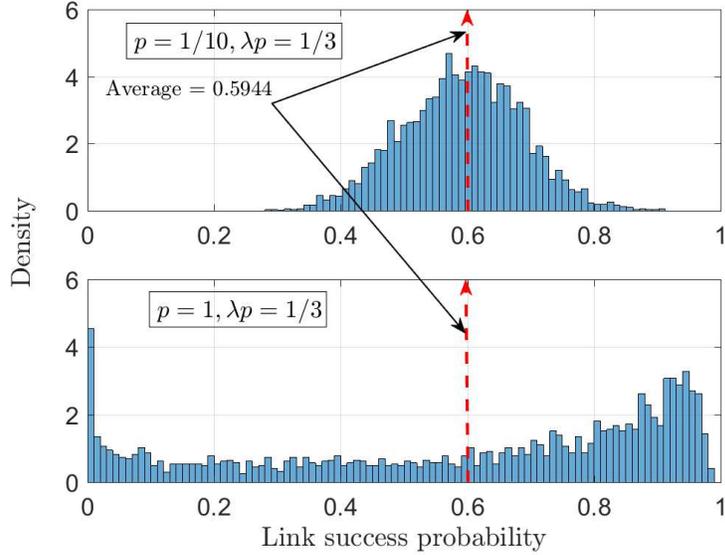}\vspace*{-1mm}
\caption{The histogram of the empirical probability density function of the link success probability in a Poisson bipolar network with ALOHA channel access scheme for transmit probabilities $p = 1/10$ and $p = 1$. Both cases have the same mean success probability of $p_{\rm{s}}(\theta) = 0.5944$, but we see a different distribution of link success probabilities for different values of the pair density $\lambda$ and transmit probability $p$. For $p = 1/10$, the link success probabilities mostly lie between $0.4$ and $0.8$ (concentrated around their mean), while for $p = 1$, they are spread much more widely. The SIR threshold $\theta = -10~\mathrm{dB}$, distance between a transmitter and its receiver $R = 1$, path loss exponent $\alpha = 4$, and $\lambda p = 1/3$.}
\label{fig:hist}\vspace*{-3mm}
\end{figure}

The conditional link success probability $P_{\rm{s}}(\theta)$ (and thus the meta distribution $\eta(\theta, \epsilon)$) allows us to directly calculate the standard (mean) success probability, which is a key quantity of interest in wireless networks. In particular, we can express the mean success probability as
\begin{equation*}
p_{\rm{s}}(\theta) \triangleq \mathbb{P}(\mathsf{SIR} > \theta) = \mathbb{E}^{!\mathrm{t}}(P_{\rm{s}}(\theta)) = \int_{0}^{1} \eta(\theta, x)\mathrm{d}x,
\end{equation*}
where the SIR is calculated at the typical receiver and $\mathbb{E}^{!\mathrm{t}}(\cdot)$ denotes the expectation with respect to the reduced Palm distribution. The standard success probability can be easily calculated by taking the average of the link success probabilities. Hence, in a realization of the network, $p_{\rm{s}}(\theta)$ can be interpreted as a spatial average which provides limited information about the outage performance of an individual link. As Fig.~\ref{fig:hist} shows, for a Poisson bipolar network with ALOHA where each transmitter has an associated receiver at a distance $R$, depending on the network parameters, the distribution of $P_{\rm{s}}(\theta)$ varies greatly for the same $p_{\rm{s}}(\theta)$. Hence the link success probability distribution is a much more comprehensive metric than the mean success probability that is usually considered. Since the SOC can be evaluated using the distribution of link success probabilities, it provides fine-grained information about the network. \vspace*{-1mm}

\subsection{Contributions}
This paper makes the following contributions:
\begin{itemize}
\item We introduce a new notion of capacity---the spatial outage capacity.
\item For the Poisson bipolar network with Rayleigh fading and ALOHA, we give exact and approximate expressions of the density of reliable links. We also derive simple upper and lower bounds on the SOC.
\item We show the trade-off between the density of active links and the fraction of reliable links.
\item In the high-reliability regime where the target outage probability is close to $0$, we give a closed-form expression of the SOC and prove that the SOC is achieved at $p = 1$. For Rayleigh distributed link distances, we show that the density of reliable links is asymptotically independent of the density of (potential) transmitters $\lambda$ as $\epsilon \to 0$.
\end{itemize}\vspace*{-1mm}

\subsection{Related Work}
For Poisson bipolar networks, the mean success probability $p_{\rm{s}}(\theta)$ is calculated in~\cite{zorzi_1995} and \cite{baccelli_2006}. For ad hoc networks modeled by the Poisson point process (PPP), the link success probability $P_{\rm{s}}(\theta)$ is studied in \cite{baccelli_infocom}, where the focus is on the mean local delay, {\em i.e.}, the $-1$st moment of $P_{\rm{s}}(\theta)$ in our notation. The notion of the \textit{transmission capacity} (TC) is introduced in~\cite{weber_2005}, which is defined as the maximum density of successful transmissions provided the outage probability {\em of the typical user} stays below a predefined threshold $\epsilon$. While the results obtained in~\cite{weber_2005} are certainly important, the TC does not represent the maximum density of successful transmissions for the target outage probability, as claimed in~\cite{weber_2005}, since the metric implicitly assumes that each link in a realization of the network is typical.

A version of the TC based on the link success probability distribution is introduced in~\cite{ganti_2010}, but it does not consider a MAC scheme, \ie, all nodes always transmit ($p = 1$). The choice of $p$ is important as it greatly affects the link success probability distribution as shown in Fig.~\ref{fig:hist}. In this paper, we consider the general case with the transmit probability $p \in (0,1]$. 

The meta distribution $\eta(\theta,\epsilon)$ for Poisson bipolar networks with ALOHA and cellular networks is calculated in~\cite{martin_meta_2016}, where a closed-form expression for the moments of $P_{\rm{s}}(\theta)$ is obtained, and an exact integral expression and simple bounds on $\eta(\theta,\epsilon)$ are provided. A key result in~\cite{martin_meta_2016} is that, for constant transmitter density $\lambda p$, as the Poisson bipolar network becomes very dense ($\lambda \to \infty$) with a very small transmit probability ($p \to 0$), the disparity among link success probabilities vanishes and all links have the same success probability, which is the mean success probability $p_{\rm{s}}(\theta)$. For the Poisson cellular network, the meta distribution of the SIR is calculated for the downlink and uplink scenarios with fractional power control in~\cite{yuanjie}, with base station cooperation in~\cite{cui_tcom}, and for D2D networks underlaying the cellular network (downlink) in~\cite{martin_d2d}. Furthermore, the meta distribution of the SIR is calculated for millimeter-wave D2D networks in~\cite{deng2017} and for D2D networks with interference cancellation in~\cite{yuanjie_letter}.\vspace*{-1mm}

\subsection{Comparison of the SOC with the TC}
The TC defined in \cite{weber_2005} can be written as 
\begin{equation*}
c(\theta, \epsilon) \triangleq (1-\epsilon)\sup\lbrace\lambda p: \mathbb{E}^{!\rm{t}}(P_{\rm{s}}(\theta)) > 1-\epsilon\rbrace, 
\end{equation*}
while the SOC can be expressed as 
\begin{equation*}
S(\theta, \epsilon) \triangleq \sup_{\lambda, p}\lbrace\lambda p \mathbb{P}(P_{\rm{s}}(\theta) > 1-\epsilon)\rbrace.
\end{equation*}
The mean success probability $p_{\rm{s}}(\theta) \triangleq \mathbb{E}^{!\rm{t}}(P_{\rm{s}}(\theta))$ depends only on the product $\lambda p$ and is monotonic. Hence the TC can be written as $c(\theta, \epsilon) \triangleq (1-\epsilon)p_{\rm{s}}^{-1}(1-\epsilon)$. The TC yields the maximum density of links such that the \textit{typical link} satisfies the outage constraint. In other words, in the TC framework, the outage constraint is applied at the typical link, {\em i.e.}, after averaging over the point process. This means that the outage constraint is not applied at the actual links, but at a fictive link whose SIR statistics correspond to the average over all links. The supremum is taken over only one parameter, namely $\lambda p$. On the other hand, in the SOC framework, the outage constraint is applied at \textit{each individual link}.\footnote{Hence the TC can be interpreted as a mean-field approximation of the SOC.} It accurately yields the maximum density of links that satisfy an outage constraint. This means that $\lambda$ and $p$ need to be considered separately. We further illustrate the difference between the SOC and the TC through the following example.
\begin{example}[\textbf{Difference between the SOC and the TC}]
\label{ex:only}
For Poisson bipolar networks with ALOHA and SIR threshold $\theta = 1/10$, link distance $R = 1$, path loss exponent $\alpha = 4$, and target outage probability $\epsilon = 1/10$, $c(1/10, 1/10) = 0.0608$ (see~\cite[(4.15)]{weber_now}), which is achieved at $\lambda p = 0.0675$. At this value of the TC, $p_{\rm{s}}(\theta) = 0.9$. But at $p = 1$, actually only $82\%$ of the active links satisfy the $10\%$ outage. Hence the density of links that achieve $10\%$ outage is only $0.055$. On the other hand, $S(1/10,1/10) = 0.09227$ which is the actual maximum density of concurrently active links that have an outage probability smaller than $10\%$. The SOC point corresponds to $\lambda = 0.23$ and $p = 1$, resulting in $p_{\rm{s}}(\theta) = 0.6984$. Thus the maximum density of links given the $10\%$ outage constraint is more than $50\%$ larger than the TC.\vspace*{-1mm} 
\end{example}

The version of the TC proposed in \cite{ganti_2010} applies an outage constraint at each link, similar to the SOC, but assumes that each link is always active (\textit{i.e.}, there is no MAC scheme) and calculates the maximum density of concurrently active links subject to the constraint that a certain fraction of active links satisfy the outage constraint. Such a constraint is not required by our definition of the SOC, and the SOC corresponds to the actual density of active links that satisfy the outage constraint.\vspace*{-1mm}

\subsection{Organization of the Paper}
The rest of the paper is organized as follows. In Sec.~\ref{sec:net_mod}, we provide the network model, formulate the SOC, give upper and lower bounds on the SOC, and obtain an exact closed-form expression of the SOC in the high-reliability regime. In Sec.~\ref{sec:rand_link}, we consider the random link distance case where the link distances are Rayleigh distributed. We draw conclusions in Sec.~\ref{sec:conclusions}.

\section{Poisson Bipolar Networks with Deterministic Link Distance} 
\label{sec:net_mod}
As seen from Def.~\ref{def:soc}, the notion of the SOC is applicable to a wide variety of wireless networks. To gain crisp insights into the design of wireless networks, in this paper, we study the SOC for Poisson bipolar networks where we consider deterministic as well as random link distances and obtain analytical results for both cases. Table~\ref{tab:notation} provides the key notation used in the paper.

\begin{table}
\caption{Summary of Notation} \label{tab:notation}
\begin{center}
\renewcommand{\arraystretch}{1.05}
\begin{tabular}{|c | p{7.1cm}| }
\hline 
 {Notation} & {\hspace{2cm}}{Definition/Meaning}
\\
\hline
\hline 

$\Phi, \Phi_{\rm t}$ & Point process of transmitters \\
$\theta$ & SIR threshold \\
$\epsilon$ & Target outage probability\\
$S(\theta, \epsilon)$ & Spatial outage capacity (SOC) \\
$\eta(\theta, \epsilon)$ & Fraction of reliable transmissions\\
$\lambda$ & Density of potential transmitters\\
$\mu$ & Density of receivers for the random link distances case\\
$p$ & Fraction of links that are active at a time\\
$\lambda_\epsilon$ & Density of reliable transmissions \\
$P_{\rm{s}}(\theta)$ & Conditional link success probability \\
$p_{\rm{s}}(\theta)$ & Mean success probability \\
$\alpha$ & Path loss exponent \\
$\delta$ & $2/\alpha$\\
$M_{b}(\theta)$ & $b$th moment of the conditional link success probability\\
$R$ & Link distance in a bipolar network\\
\hline 
\end{tabular}
\end{center}\vspace*{-4mm}
\end{table}

\subsection{Network Model}
We consider the Poisson bipolar network model in which the locations of transmitters form a homogeneous Poisson point process (PPP) $\Phi \subset \mathbb{R}^2$ with density $\lambda$~\cite[Def. 5.8]{martin_book}. Each transmitter has a dedicated receiver at a distance $R$ in a uniformly random direction. In a time slot, each node in $\Phi$ independently transmits at unit power with probability $p$ and stays silent with probability $1-p$. Thus the active transmitters form a homogeneous PPP with density $\lambda p$. We consider a standard power law path loss model with path loss exponent $\alpha$. We assume that a channel is subject to independent Rayleigh fading with channel power gains as i.i.d. exponential random variables with mean $1$. 

We focus on the interference-limited case, where the received SIR is a key quantity of interest. To the PPP, we add a (desired) transmitter at location ($R,0$) and a receiver at the origin $o$. Under the expectation over the PPP, this link is the typical link. The success probability $p_{\rm{s}}(\theta)$ of the typical link is the ccdf of the SIR calculated at the origin. For Rayleigh fading, from \cite{baccelli_2006,martin_book}, it is known that
\begin{equation}
p_{\rm{s}}(\theta) = \exp\left(-\lambda p C \theta^{\delta}\right),
\label{eq:mp_s}
\end{equation}
where $C \triangleq  \pi R^2 \Gamma(1+\delta)\Gamma(1-\delta)$ with $\delta \triangleq 2/\alpha$. The model is scale-invariant in the following sense: The SIR of all links in any realization of the bipolar network with transmitter locations $\varphi$ remains unchanged if the plane is scaled by an arbitrary factor $a>0$. Such scaling results in transmitter locations $a\varphi$ and link distances $aR$. The density of the scaled network is $\lambda/a^2$. By setting $a=1/R$ to obtain unit distance links, the resulting density is $\lambda R^2$. Hence without loss of generality, we can set $R=1$. Applied to the meta distribution and the SOC, this means that the model with parameters $(R,\lambda)$ behaves exactly the same as the model with parameters $(1,\lambda R^2)$.\vspace*{-1mm}

\subsection{Exact Formulation}

Observe from Def.~\ref{def:soc} that the SOC depends on $\eta(\theta,\epsilon) = \mathbb{P}(P_{\rm{s}}(\theta) > 1-\epsilon)$ whose direct calculation seems infeasible. But the moments of $P_{\rm{s}}(\theta)$ are available in closed-form~\cite{martin_meta_2016}, from which we can derive exact and approximate expressions of $\lambda_{\epsilon}$ and obtain simple upper and lower bounds on the SOC. Let $M_b(\theta)$ denote the $b$th moment of $P_{\rm{s}}(\theta)$, {\em i.e.,}
\begin{equation}
M_b(\theta) \triangleq \mathbb{E}\left(P_{\rm{s}}(\theta)^{b}\right).
\label{eq:moments_Ps}
\end{equation}
The mean success probability is $p_{\rm{s}}(\theta) \equiv M_1(\theta)$. 

From~\cite[Thm. 1]{martin_meta_2016}, we can express $M_b(\theta)$ as
\begin{equation}
M_b(\theta) = \exp\left(-\lambda C \theta^{\delta} D_b(p,\delta)\right), \quad b\in\mathbb{C},
\label{eq:moment1}
\end{equation}
where
\begin{equation}
D_b(p,\delta)\triangleq \sum_{k=1}^\infty \binom bk\binom{\delta-1}{k-1} p^k, \quad p,\delta\in (0,1].
\label{eq:Db}
\end{equation}
For $b \in \mathbb{N}$, the sum is finite and $D_b(p,\delta)$ becomes a polynomial which is termed \textit{diversity polynomial} in~\cite{martin_diversity}. The series in (7) converges for $p < 1$, and at $p = 1$ it is defined if $b \notin \mathbb{Z}^{-}$ or $b + \delta \notin \mathbb{Z}^{-}$ and converges if $\Re(b+\delta) >0$. Here $\Re(z)$ is the real part of the complex number $z$. For $b = 1$ (the first moment), $D_1(p,\delta) = p$, and we get the expression of $p_{\rm{s}}(\theta)$ as in \eqref{eq:mp_s}. We can also express $D_b(p,\delta)$ using the Gaussian hypergeometric function $_2F_1$ as
\begin{equation}
 D_b(p,\delta)=pb\:_2F_1(1-b, 1-\delta; 2; p).
 \label{eq:db_hyper}
\end{equation}
Using the Gil-Pelaez theorem~\cite{gp_theorem}, the exact expression of $\lambda_{\epsilon} = \lambda p \eta(\theta, \epsilon)$ can be obtained in an integral form from that of $\eta(\theta,\epsilon)$ given in~\cite[Cor. 3]{martin_meta_2016} as
\begin{equation}
\lambda_{\epsilon} =  \frac{\lambda p}{2}-\frac{\lambda p}{\pi} \int\limits_0^\infty \frac{\sin(u\ln (1-\epsilon)+ \lambda C\theta^\delta\Im(D_{ju}))}{ue^{\lambda C\theta^\delta \Re(D_{ju})}}\mathrm{d}u ,
\label{eq:exact_soc}
\end{equation}
where $j \triangleq \sqrt{-1}$, $D_{ju}=D_{ju}(p,\delta)$, and $\Im(z)$ is the imaginary part of the complex number $z$. The SOC is then obtained by taking the supremum of $\lambda_{\epsilon}$ over $\lambda > 0$ and $p \in (0,1]$.\vspace*{-1mm}
\subsection{Approximation with Beta Distribution}

We can accurately approximate $\lambda_{\epsilon}$ in a semi-closed form using the beta distribution, as shown in~\cite{martin_meta_2016}. The rationale behind such approximation is that the support of the link success probability $P_{\rm{s}}(\theta)$ is $[0,1]$, making the beta distribution a natural choice. With the beta distribution approximation, $\lambda_{\epsilon}$ can be approximated as
\begin{equation}
\lambda_{\epsilon} \approx \lambda p \left(1 - I_{\epsilon}\left(\frac{\mu \beta}{1-\mu},\beta\right)\right),
\label{eq:tau_app}
\end{equation}
where $I_\epsilon(y,z) \triangleq \int_{0}^{1-\epsilon}t^{y -1}(1-t)^{z - 1}\mathrm{d}t/B(y,z)$ is the regularized incomplete beta function with $B(\cdot,\cdot)$ denoting beta function, $\mu = M_1$, and $\beta = (M_1 -M_2)(1-M_1)/(M_2 - M_1^2)$.

The advantage of the beta approximation is the faster computation of $\lambda_{\epsilon}$ compared to the exact expression without losing much accuracy~\cite[Tab. I, Fig. 4]{martin_meta_2016} (also see Fig.~\ref{fig:asym_eps_0} of this paper). In general, it is difficult to obtain the SOC analytically due to the forms of $\lambda_{\epsilon}$ given in \eqref{eq:exact_soc} and \eqref{eq:tau_app}. But we can obtain the SOC numerically with ease. We can also gain useful insights considering some specific scenarios, on which we focus in the following subsection.\vspace*{-1mm}

\subsection{Constrained SOC}
\subsubsection{Constant $\lambda p$ in dense networks}
For constant $\lambda p$ (or, equivalently, a fixed $p_{\rm{s}}(\theta)$), we now study how the density of reliable links $\lambda_{\epsilon}$ behaves in an ultra-dense network. Given $\theta$, $\alpha$, and $\epsilon$, this case is equivalent to asking how $\lambda_{\epsilon}$ varies as $\lambda \to \infty$ while letting $p \to 0$ for constant transmitter density $\lambda p$ (constant $p_{\rm{s}}(\theta)$). We denote the constrained SOC by $\tilde{S}(\theta, \epsilon)$.
\begin{lemma}[\textbf{$\boldsymbol{p \to 0}$ for constant $\boldsymbol{\lambda p}$}] \label{th:p_to_0} Let $\nu = \lambda p$. Then, for constant $\nu$ while letting $p \to 0$, the SOC constrained on the density of concurrent transmissions is
\begin{equation}
\label{eq:binary1}
\tilde{S}(\theta,\epsilon) = \left\{
  \begin{array}{l l}
    \lambda p , & \quad \mathrm{if}~ 1 -\epsilon < p_{\rm{s}}(\theta) \\
    0, & \quad \mathrm{if}~ 1 - \epsilon > p_{\rm{s}}(\theta).\\
  \end{array} \right.
\end{equation}
\end{lemma}
\begin{IEEEproof}
Applying Chebyshev's inequality to \eqref{eq:meta_dist}, for $1-\epsilon < p_{\rm{s}}(\theta) = M_1$, we have
\begin{equation}
\eta(\theta,\epsilon) > 1 - \frac{\mathrm{var}(P_{\rm{s}}(\theta))}{((1-\epsilon)-M_1)^2},
\label{eq:cheb_1} 
\end{equation}
where $\mathrm{var}(P_{\rm{s}}(\theta)) = M_2 - M_1^2$ is the variance of $P_{\rm{s}}(\theta)$. From~\cite[Cor. 1]{martin_meta_2016}, for constant $\nu$, we know that $\lim\limits_{\underset{\lambda p = \nu}{p \to 0}}\mathrm{var}(P_{\rm{s}}(\theta)) = 0.$
Hence the lower bound in \eqref{eq:cheb_1} approaches $1$, which leads to $\eta(\theta,\epsilon) \to 1$. This results in the SOC constrained on the density of concurrent transmissions equal to $\lambda p$.

On the other hand, for $1-\epsilon > M_1$, 
\begin{equation}
\eta(\theta,\epsilon) \leq \frac{\mathrm{var}(P_{\rm{s}}(\theta))}{((1-\epsilon)-M_1)^2}.
\label{eq:cheb_2} 
\end{equation}
As we let $p \to 0$ for constant $\nu$, the upper bound in \eqref{eq:cheb_2} approaches $0$, which leads to $\eta(\theta,\epsilon) \to 0$. This results in the SOC constrained on the density of concurrent transmissions equal to $0$.
\end{IEEEproof}
In fact, as $\mathrm{var}(P_{\rm{s}}(\theta)) \to 0$, the ccdf of $P_{\rm{s}}(\theta)$ approaches a step function that drops from $1$ to $0$ at the mean of $P_{\rm{s}}(\theta)$, \textit{i.e.}, at $1-\epsilon = p_{\rm{s}}(\theta)$. This behavior is in agreement with \eqref{eq:binary1}.

\textbf{Remark}: Lemma~\ref{th:p_to_0} shows that, if $p\to 0$ while $\lambda p$ is fixed to the value $\nu$ at which $p_{\rm{s}}(\theta)$ equals the target reliability $1-\epsilon$, the maximum value of the constrained SOC is the value of the TC times $1/(1-\epsilon)$, and that value of the TC is $\nu(1-\epsilon)$.

This observation can be explained as follows: As $p\to 0$ while keeping $\lambda p = \nu$, all links in a realization of the network have the same success probability, and that value of the success probability equals $p_{\rm{s}}(\theta)$ (\textit{i.e.}, the success probability of transmissions over the typical link)~\cite{martin_meta_2016}. This implies that, from the outage perspective, each link in the network can now be treated as if it were the typical link, as in the TC framework. If $\nu$ is initially set to a value that results in $p_{\rm{s}}(\theta) > 1-\epsilon$, we can always increase it till $p_{\rm{s}}(\theta) = 1-\epsilon$ while all active links satisfying the outage constraint (or, equivalently, the typical link satisfying the outage constraint with probability one). Accordingly, the value of the TC equals $1-\epsilon$ times the value of $\nu$ at which $p_{\rm{s}}(\theta) = 1-\epsilon$.

\begin{figure}
\centering
\includegraphics[scale=0.58]{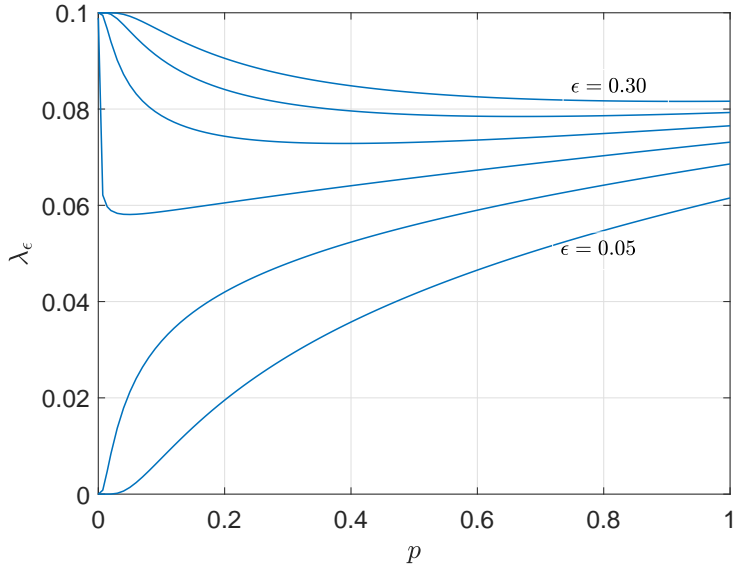}\vspace*{-2mm}
\caption{The density of reliable links $\lambda_{\epsilon}$ against the transmit probability $p$ for $\lambda p = 1/10$, $\theta = -10~\mathrm{dB}$, $\alpha = 4$, and $R = 1$. The values at the curves are $\epsilon = 0.05, 0.1, 0.15, 0.2, 0.25, 0.3$ (bottom to top). The mean success probability is $p_{\rm{s}}(\theta) = 0.855$.}
\label{fig:eff_p_lambda_eps}\vspace*{-3mm}
\end{figure}

Fig.~\ref{fig:eff_p_lambda_eps} shows that at small values of the target outage probability $\epsilon$, the density of reliable transmissions monotonically increases with $p$. On the other hand, at larger values of $\epsilon$, it first decreases with $p$.

\subsubsection{$\lambda p \to 0$}
For $\lambda p \to 0$, $\lambda_{\epsilon}$ depends linearly on $\lambda p$, which we prove in the next lemma.
\begin{lemma}[$\boldsymbol{\lambda_{\epsilon}}$ \textbf{as} $\boldsymbol{\lambda p \to 0}$]\label{th:lam_0}
As $\lambda p \to 0$,
\begin{align*}
\lambda_{\epsilon} \sim \lambda p.
\end{align*}
\end{lemma}
\begin{IEEEproof}
As $\lambda p \to 0$, $M_1$ approaches $1$ and thus  $\mathrm{var}(P_{\rm{s}}(\theta)) = M_1^2(M_1^{p(\delta -1)} - 1)$ approaches $0$. Since $\epsilon \in (0,1)$, we have $1-\epsilon < M_1$ as $\lambda p \to 0$. Using Chebyshev's inequality for $1-\epsilon < M_1$ as in \eqref{eq:cheb_1} and letting $\mathrm{var}(P_{\rm{s}}(\theta)) \to 0$, the lower bound in \eqref{eq:cheb_1} approaches $1$, leading to $\eta(\theta,\epsilon) \to 1$. 
\end{IEEEproof}
%The supremum of $\lambda_{\epsilon}$ can be achieved by optimizing $\lambda_{\epsilon}$ over $\lambda$, i.e., $S\sim \sup\limits_{\lambda} \lambda p$ as $\lambda \to 0$.
Lemma~\ref{th:lam_0} can be understood as follows. As $\lambda p \to 0$, the density of active transmitters is very small. Thus each transmission succeeds with high probability and $\eta(\theta,\epsilon) \to 1$. In this regime, the density of reliable links $\lambda_{\epsilon}$ is directly given by $\lambda p$.

The case $\lambda p \to 0$ can be interpreted in two ways: 1) $\lambda \to 0$ for constant $p$ and 2) $p \to 0$ for constant $\lambda$. Lemma~\ref{th:lam_0} is valid for both cases, or any combination thereof. The case of constant $p$ is relevant since it can be interpreted as a delay constraint: As $p$ gets smaller, the probability that a node makes a transmission attempt in a slot is reduced, increasing the delay. Since the mean delay until successful reception is larger than the mean channel access delay $1/p$, it gets large for small values of $p$. Thus, a delay constraint prohibits $p$ from getting too small.

\begin{figure}
\centering
\includegraphics[scale=0.58]{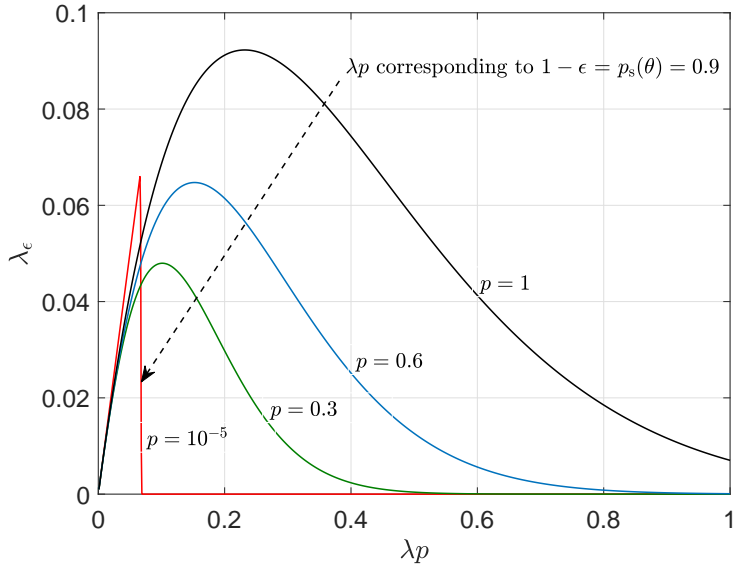}\vspace*{-2mm}
\caption{The density of reliable links $\lambda_{\epsilon}$ given in \eqref{eq:ss1} for different values of the transmit probability $p$ for $\theta = -10~\mathrm{dB}$, $\alpha = 4$, and $\epsilon = 1/10$. Observe that the slope of $\lambda_{\epsilon}$ is one for small $\lambda p$. The dashed arrow points to the value of $\lambda p = 0.0675$, which corresponds to $1-\epsilon = p_{\rm{s}}(\theta) = 0.9$.}
\label{fig:lam_prod_soc}\vspace*{-3mm}
\end{figure}

Fig.~\ref{fig:lam_prod_soc} illustrates Lemma~\ref{th:lam_0}. Also, observe that, as $p \to 0$ ($p = 10^{-5}$ in Fig.~\ref{fig:lam_prod_soc}), $\lambda_{\epsilon}$ increases linearly with $\lambda p$ until $\lambda p$ reaches the value $0.0675$ which corresponds to $p_{\rm{s}}(\theta)= 1-\epsilon = 0.9$ and then drops to $0$. This behavior is in accordance with Lemma~\ref{th:p_to_0}. In general, as $\lambda p$ increases, $\lambda_{\epsilon}$ increases first and then decreases after a tipping point. This is due to the two opposite effects of $\lambda p $ on $\lambda_{\epsilon}$: The density $\lambda p$ of active transmitters increases, but at the same time, more active transmissions cause higher interference, which in turn, reduces the fraction $\eta(\theta,\epsilon)$ of links that have a reliability at least $1-\epsilon$. 

\begin{figure*} [t]
    \centering
  \subfloat[]{ \includegraphics[width=0.49\linewidth]{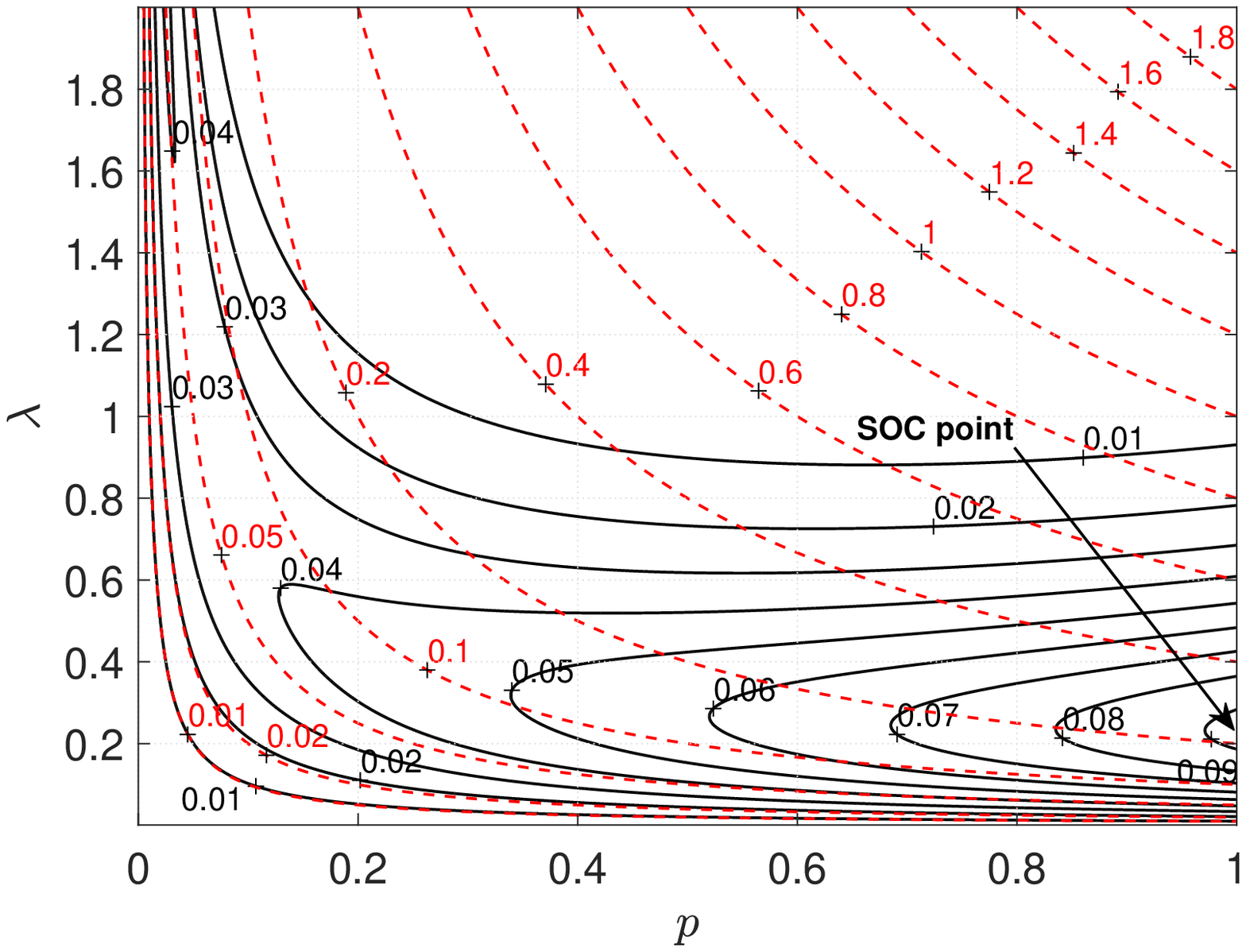}\label{fig:lam_prod_soc1}}
   \hfill
  \subfloat[]{
        \includegraphics[width=0.49\linewidth]{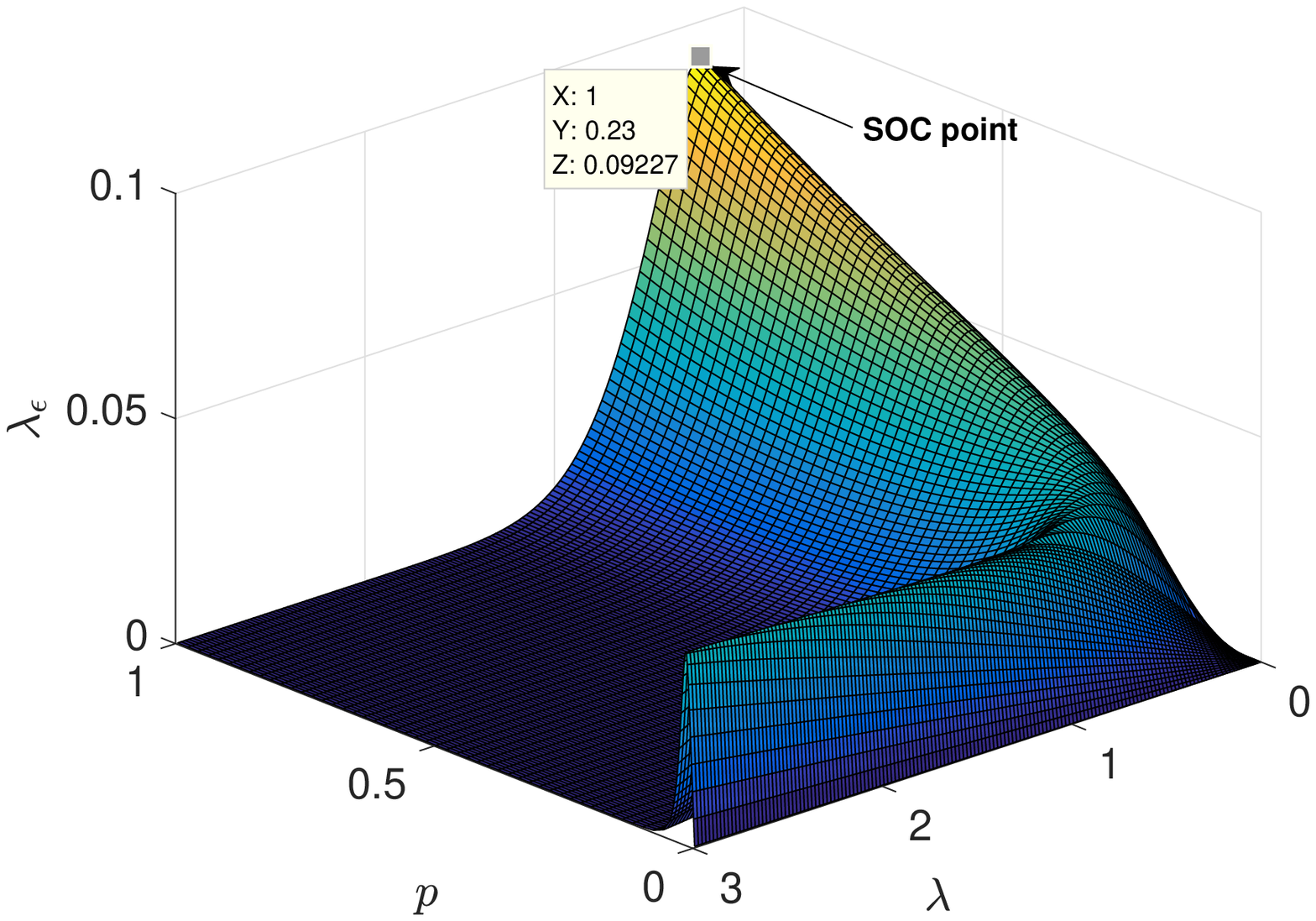}\label{fig:3d_lam_p_tau}}
    
  \caption{(a) Contour plots of $\lambda_{\epsilon}$ and the product $\lambda p$ for $\theta = -10~\mathrm{dB}$, $\alpha = 4$, and $\epsilon = 1/10$. The solid lines represent the contour curves for $\lambda_{\epsilon}$ and the dashed lines represent the contour curves for $\lambda p$. The numbers in ``\textcolor{black}{black}'' and ``\textcolor{red}{red}'' indicate the contour levels for $\lambda_{\epsilon}$ and $\lambda p$, respectively. The SOC is $S(\theta,\epsilon) = 0.09227$. The values of $\lambda$ and $p$ at the SOC point are $0.23$ and $1$, respectively, and the corresponding mean success probability is $p_{\rm{s}}(\theta) = 0.6984$. The arrow corresponding to the ``SOC point'' points to the pair of $\lambda$ and $p$ for which the SOC is achieved. (b) Three-dimensional plot of $\lambda_{\epsilon}$ corresponding to the contour plot.}\vspace*{-3mm}
\end{figure*}

The contour plot in Fig.~\subref*{fig:lam_prod_soc1} visualizes the trade-off between $\lambda p$ and $\eta(\theta,\epsilon)$. The contour curves for small values of $\lambda p$ run nearly parallel to those for $\lambda_{\epsilon}$, indicating that $\eta(\theta,\epsilon)$ is close to $1$. Specifically, the contour curves for $\lambda p = 0.01$ and  $\lambda p = 0.02$ match those for $\lambda_{\epsilon} = 0.01$ and $\lambda_{\epsilon} = 0.02$ almost exactly. This behavior is in accordance with Lemma~\ref{th:lam_0}. In contrast, for large values of $\lambda p$, the decrease in $\eta(\theta,\epsilon)$ dominates $\lambda_{\epsilon}$. Also, for larger values of $\lambda$ ($\lambda > 0.4$ for Fig.~\subref*{fig:lam_prod_soc1}), $\lambda_{\epsilon}$ first increases and then decreases with the increase in $p$. This behavior is due to the following trade-off in $p$. For a small $p$, there are few active transmitters in the network per unit area, but a higher fraction of links are reliable. On the other hand, a large $p$ means more active transmitters per unit area, but also a higher interference which reduces the fraction of reliable links. For $\lambda < 0.4$, the increase in the density of active transmitters dominates the decrease in $\eta(\theta,\epsilon)$, and $\lambda_{\epsilon}$ increases monotonically with $p$. The three-dimensional plot corresponding to the contour plot in Fig.~\subref*{fig:lam_prod_soc1} is shown in Fig.~\subref*{fig:3d_lam_p_tau}.
\vspace*{-1mm}
\subsection{Bounds on the SOC}
\label{sec:det_up}
In this subsection, we obtain simple upper and lower bounds on the SOC. 

\begin{theorem}[\textbf{Upper bound on the SOC}]
\label{thm:soc_up}
For any $b > 0$, the SOC is upper bounded as
\begin{equation}
\label{eq:binary3}
 S(\theta,\epsilon) \leq \left\{
  \begin{array}{l l}
    \frac{1}{e\pi \theta^{\delta} \Gamma(1-\delta)\Gamma(1+\delta)} \frac{1}{b(1-\epsilon)^{b}}, & 0 < b \leq  1,\\\frac{1}{e\pi  \theta^{\delta} \Gamma(1-\delta)} \frac{\Gamma(b)}{\Gamma(b+\delta)(1-\epsilon)^{b}}, & b > 1.    
  \end{array} \right.
\end{equation}
\end{theorem}
\begin{IEEEproof}
Using Markov's inequality, $\eta(\theta,\epsilon)$ can be upper bounded as
\begin{align}
\eta(\theta,\epsilon) \leq \frac{M_b(\theta)}{(1-\epsilon)^b}, \quad b > 0,
\label{eq:mark}
\end{align}
where $M_b(\theta) = e^{-\lambda C \theta^{\delta}D_b(p,\delta)}$. Hence we can upper bound the SOC as
\begin{align*}
S(\theta, \epsilon) \leq  S_{\rm{u}},
\end{align*}
where 
\begin{align}
S_{\rm{u}} =  \sup_{\lambda, p}\lambda p \frac{e^{-\lambda C \theta^{\delta}D_b(p,\delta)}}{(1-\epsilon)^{b}},
\label{eq:bound_tau1}
\end{align}
with $C = \pi \Gamma(1-\delta)\Gamma(1+\delta)$ and $D_b(p,
\delta) = pb\:_2F_1(1-b, 1-\delta; 2; p)$. Let us denote $f_{\rm{u}}(\lambda, p) = \lambda p e^{-\lambda C \theta^{\delta}D_b(p,\delta)}$. We can then write 
\begin{align*}
\frac{\partial f_{\rm{u}}(\lambda, p)}{\partial \lambda} = \underbrace{p e^{-\lambda C \theta^{\delta}D_b(p,\delta)}}_{> 0}(1 - \lambda C \theta^{\delta}D_b(p,\delta)).
\end{align*}
Setting $\frac{\partial f_{\rm{u}}(\lambda, p)}{\partial \lambda} = 0$, we obtain the critical point as
$\lambda_0(p) = 1/(C \theta^{\delta}D_b(p,\delta))$. For any given $p$, the objective function is quasiconcave. Thus $\lambda_0(p)$ is the global optimum for each $p$. As a result, the optimization problem in~\eqref{eq:bound_tau1} reduces to
\begin{align}
S_{\rm{u}} &= \frac{1}{(1-\epsilon)^b}\sup_{p} f_{\rm{u}}(\lambda_0(p), p) \nonumber \\
&= \frac{1}{eC \theta^{\delta}b(1-\epsilon)^{b}}~\sup\limits_{p} ~\frac{1}{\:_2F_1(1-b, 1-\delta; 2; p)}.
\label{eq:sub33}
\end{align}
For $0 <b < 1$, $\:_2F_1(1-b, 1-\delta; 2; p)$ monotonically increases with $p$. In this case, $p \to 0$ solves \eqref{eq:sub33}. 
%Since $\:_2F_1(1-b, 1-\delta; 2; 0) = 1$,
%\begin{align}
%S_{\rm{u}} = \frac{1}{e\pi \theta^{\delta} \Gamma(1-\delta)\Gamma(1+\delta)} \frac{1}{b(1-\epsilon)^{b}}, \quad 0 < b < 1.
%\label{eq:socu1}
%\end{align}
On the other hand, for $b > 1$, $\:_2F_1(1-b, 1-\delta; 2; p)$ monotonically decreases  with $p$. Thus $p = 1$ solves \eqref{eq:sub33}. Overall the value of $p$ that solves \eqref{eq:sub33} is
\begin{align}
 p_0  \left\{
  \begin{array}{l l}
    \to 0, & 0 < b < 1,\\=1, & b > 1.    
  \end{array} \right.
  \label{eq:sub22}
\end{align}
Note that the objective function in \eqref{eq:sub33} is monotonic in $p$. Hence $p_0$ in \eqref{eq:sub22} is again the global optimum.

Finally, for $0 < b < 1$, the upper bound on the SOC is obtained by substituting $p = 0$ in the objective of \eqref{eq:sub33}. Since $\:_2F_1(1-b, 1-\delta; 2; 0) = 1$,
\begin{align}
S_{\rm{u}} = \frac{1}{e\pi \theta^{\delta} \Gamma(1-\delta)\Gamma(1+\delta)} \frac{1}{b(1-\epsilon)^{b}}, \quad 0 < b < 1.
\label{eq:socu1}
\end{align}
Similarly since $b\:_2F_1(1-b, 1-\delta; 2; 1) = \frac{\Gamma(b+\delta)}{\Gamma(b)\Gamma(1+\delta)}$,
\begin{align}
S_{\rm{u}} = \frac{1}{e\pi \theta^{\delta} \Gamma(1-\delta)} \frac{\Gamma(b)}{\Gamma(b+\delta)(1-\epsilon)^{b}}, \quad b > 1.
\label{eq:socu2}
\end{align}
For $b = 1$, the hypergeometric function returns $1$ irrespective of the other parameters, and thus \eqref{eq:socu1} and \eqref{eq:socu2} are identical.
\end{IEEEproof}
The tightest Markov upper bound can be obtained by minimizing $S_{\rm{u}}$ in \eqref{eq:socu1} and \eqref{eq:socu2} over $b$. Now, the value of $b$ that minimizes $S_{\rm{u}}$ in \eqref{eq:socu1}  is
\begin{align}
b_{\rm{m}} = -\frac{1}{\ln(1-\epsilon)}. 
\label{eq:bm_small}
\end{align}
Since $S_{\rm{u}}$ takes two different values depending on whether $0 < b \leq1$ or $b >1$, to obtain the tightest Markov upper bound, we need to consider following two cases based on whether $b_{\rm{m}} \in (0,1]$ or  $b_{\rm{m}} > 1$.

\textit{1) $\boldsymbol{b_{\rm{m}} \in (0,1]}$}: If $b_{\rm{m}} \in (0,1]$, it is the optimum value of $b$ that minimizes $S_{\rm{u}}$ since $S_{\rm{u}}$ in \eqref{eq:socu1} is smaller than $S_{\rm{u}}$ in \eqref{eq:socu2} for $0 < b < 1$, greater for $b > 1$, and equal to for $b = 1$. From \eqref{eq:bm_small}, it is apparent that the case $b_{\rm{m}} \in (0,1]$ is equivalent to $\epsilon \in [0.6321,1)$. Hence, if $\epsilon \in [0.6321,1)$, the optimum $b$ that gives the tightest Markov upper bound is given by \eqref{eq:bm_small}. Substituting $b = b_{\rm{m}}$ in \eqref{eq:socu1}, we get the exact closed-form expression of the tightest Markov upper bound as 
\begin{align}
S_{\rm{u}}^{\rm{t}} = \frac{-\ln(1-\epsilon)}{\pi  \theta^{\delta} \Gamma(1-\delta)\Gamma(1+\delta)}, \quad \text{if}~\epsilon \in [0.6321,1),
\label{eq:soc_up_fin_ex}
\end{align}
where `$\rm{t}$' in the superscript of $S_{\rm{u}}^{\rm{t}}$ indicates the tightest bound.

\begin{figure}
\centering
\includegraphics[scale=0.58]{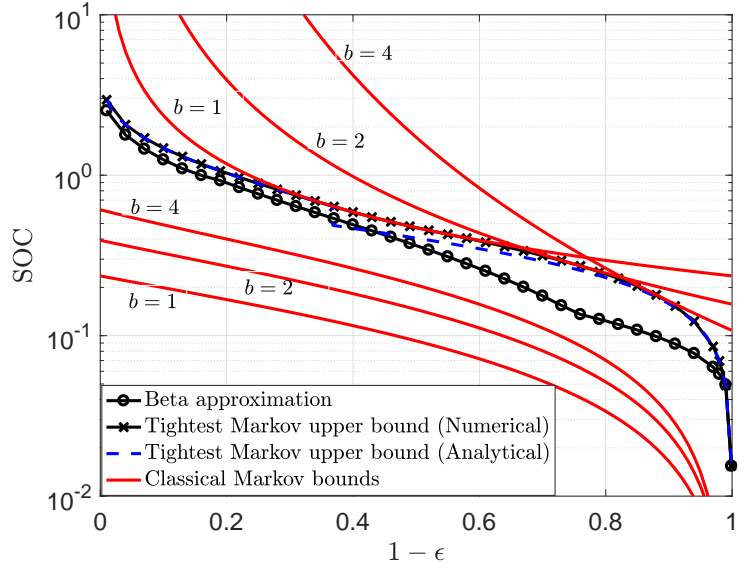}\vspace*{-1mm}
\caption{Analytical and numerical results for the SOC. The tightest Markov upper bound on the SOC obtained numerically uses \eqref{eq:socu1} and \eqref{eq:socu2}, which are optimized over $b$. The tightest Markov upper bound obtained analytically uses \eqref{eq:soc_up_fin_ex} and \eqref{eq:soc_up_fin}. The SOC upper bound obtained analytically is quite close to that obtained numerically for the almost complete range of reliability threshold $1-\epsilon$, except near $1-\epsilon = 0.3679$ (which is due to the approximation in \eqref{eq:approx_gamma}). The classical Markov bounds are plotted using \eqref{eq:socu1} and \eqref{eq:socu2} for $b =1$, $b = 2$, and $b = 4$. The lower bound for $b = 1$ is plotted using \eqref{eq:sl_main}, while the lower bounds for $b = 2$ and $b = 4$ are plotted numerically using \eqref{eq:sl1}. $\theta = -10$ dB and $\alpha = 4$.}
\label{fig:upper_det_link1}\vspace*{-3mm}
\end{figure}

2) $\boldsymbol{b_{\rm{m}} > 1}$: If $b_{\rm{m}} > 1$, \textit{i.e.}, $\epsilon \in (0, 0.6321)$, the optimum $b$ is the value of $b$ that minimizes $S_{\rm{u}}$ in \eqref{eq:socu2}. However, due to the form of $S_{\rm{u}}$ in \eqref{eq:socu2}, the optimum $b$ cannot be expressed in a closed-form. Hence the tightest Markov upper bound also cannot be expressed in a closed-form, but it can be easily evaluated numerically. Furthermore, for $b > 1$, we can get a closed-form expression of the approximate tightest Markov bound by using the approximation
\begin{align}
\frac{\Gamma(b+\delta)}{\Gamma(b)} \approx b^{\delta}
\label{eq:approx_gamma}
\end{align}
in \eqref{eq:socu2}. Then, for $b > 1$, we can express \eqref{eq:socu2} as
\begin{align}
S_{\rm{u}} \approx \frac{1}{e\pi  \theta^{\delta} \Gamma(1-\delta)} \frac{1}{b^{\delta}(1-\epsilon)^{b}}.
\label{eq:socu3}
\end{align}
The value of $b$ that minimizes \eqref{eq:socu3} is given as
\begin{equation*}
\bar{b}_{\rm{m}} =  -\frac{\delta}{\ln(1-\epsilon)}.
\end{equation*}
The corresponding closed-form expression of the tightest approximate Markov upper bound is obtained by substituting $\bar{b}_{\rm{m}}$ in \eqref{eq:socu3} which is given as
\begin{equation}
S_{\rm{u}}^{\rm{t}} \approx \left(\frac{-\ln(1-\epsilon)}{\theta}\right)^{\delta}\frac{e^{-(1-\delta)}}{\pi  \delta^{\delta} \Gamma(1-\delta)}, \quad \text{if}~\epsilon \in (0, 0.6321).
\label{eq:soc_up_fin}
\end{equation}
Fig.~\ref{fig:upper_det_link1} illustrates upper bounds on the SOC.
%We can also obtain the bound in \eqref{eq:soc_up_fin} directly by using the approximation $D_b(p,\delta) \approx p^{\delta}b^{\delta}/\Gamma(1+\delta)$ in Markov's inequality given by \eqref{eq:mark} and then optimizing over $b$, $\lambda$, and $p$. But, as Fig.~\ref{fig:upper_det_link1} shows, the bound obtained in this way is not strictly an upper bound at lower values of $1-\epsilon$, because the optimum $b$ that gives the tightest Markov upper bound lies between $0$ and $1$ at lower values of $1-\epsilon$ where \eqref{eq:soc_up_fin_ex} gives the exact tightest Markov upper bound.}

Letting $\epsilon \to 0$, from \eqref{eq:soc_up_fin}, we observe that the lower tail of the SOC decreases exponentially, \textit{i.e.}, 
\begin{align}
S(\theta, \epsilon) \lessapprox \left(\frac{\epsilon}{\theta}\right)^{\delta}\frac{e^{-(1-\delta)}}{\pi  \delta^{\delta} \Gamma(1-\delta)}, \quad \epsilon \to 0,
\label{eq:S_up_asym}
\end{align}
where `$\lessapprox$' denotes an upper bound which gets tighter asymptotically (here as $\epsilon \to 0$).
In the next subsection, we shall show that the bound in \eqref{eq:S_up_asym} is in fact asymptotically tight, \textit{i.e.}, \eqref{eq:S_up_asym} matches the exact expression of the SOC as $\epsilon \to 0$.

%where `$\lesssim$' denotes an upper bound with asymptotic equality (here as $\epsilon \to 0$).

We now obtain lower bounds on the SOC. 
\begin{theorem}[\textbf{Lower bound on the SOC}]
\label{th:low_det}
The SOC is lower bounded as
\begin{equation}
S(\theta, \epsilon) > \left(\frac{1-W((1-\epsilon)e)}{\pi\theta^{\delta}\Gamma(1+\delta)\Gamma(1-\delta)}\right)\left(\frac{e^{-(1-W((1-\epsilon)e))}-(1-\epsilon)}{\epsilon}\right), 
\label{eq:sl_main}
\end{equation}
where $W(\cdot)$ denotes the Lambert $W$ function.
\end{theorem}
\begin{IEEEproof}
By the reverse Markov's inequality, we have 
\begin{equation*}
1- \frac{\mathbb{E}^{!\rm{t}}((1-P_{\rm{s}}(\theta))^{b})}{\epsilon^{b}} < \eta(\theta,\epsilon), \quad b > 0.
\end{equation*}
For $b \in \mathbb{N}$ we can lower bound the SOC as
\begin{equation*}
S(\theta, \epsilon) > S_{\rm{l}},
\end{equation*}
where 
\begin{align}
S_{\rm{l}} = \sup_{\lambda, p} \lambda p \left(1 - \frac{\sum_{k = 0}^{b}\binom{b}{k}(-1)^{k}M_k(\theta)}{\epsilon^b}\right),
\label{eq:sl1}
\end{align}
with $M_{k}(\theta) = e^{-\lambda C \theta^{\delta}D_k(p,\delta)}$. For $b = 1$, \eqref{eq:sl1} reduces to
\begin{align}
S_{\rm{l}} = \sup_{\lambda, p} ~\lambda p\left(1 - \frac{1-e^{-\lambda  p C \theta^{\delta}}}{\epsilon}\right).
\end{align}
Since $\lambda$ and $p$ appear together as their product $\lambda p$, $S_{\rm{l}}$ can be obtained by taking the supremum over $t = \lambda p$, \textit{i.e.},
\begin{align}
S_{\rm{l}} = \sup_{t}~ \underbrace{t\left(1 - \frac{1-e^{-t C \theta^{\delta}}}{\epsilon}\right)}_{f(t)}.
\end{align}
Substituting the value of $t$ that results in $\frac{\partial f(t)}{\partial t} = 0$ in $f(t)$, we get the desired expression in \eqref{eq:sl_main}.
\end{IEEEproof}
For the values of $b \in \mathbb{R}^{+}\setminus\lbrace1\rbrace $, an analytical expression for $S_{\rm{l}}$ is difficult to obtain due to the form of \eqref{eq:sl1}, but we can easily obtain corresponding lower bounds numerically. Fig.~\ref{fig:upper_det_link1} shows Markov lower bounds on the SOC for $b = 1$, $b = 2$, and $b = 4$.\vspace*{-1mm}

\subsection{High-reliability Regime}
\label{sec:high_det}

In this section, we investigate the behavior of $\lambda_{\epsilon}$ and the SOC in the high-reliability regime, \textit{i.e.}, as $\epsilon \to 0$. To this end, we first provide an asymptote of $D_b(p,\delta)$ as $b \to \infty$, which will be used to obtain a closed-form expression of the SOC in the high-reliability regime. Then we state a simplified version of de Bruijn's Tauberian theorem (see~\cite[Thm.~4.12.9]{bingham_1987}) which allows a convenient formulation of $\eta (\theta, \epsilon) = \mathbb{P}(P_{\rm{s}}(\theta) > 1-\epsilon)$ in terms of the Laplace transform as $\epsilon \to 0$. `$\lesssim$' denotes an upper bound with asymptotic equality (here as $b \to \infty$).
\begin{lemma}[\textbf{Asymptote of} $\boldsymbol{D_b(p,\delta)}$ \textbf{as} $\boldsymbol{b \to \infty}$]
\label{lem:db_asym}
For $b \in \mathbb{R}$, we have
\begin{equation}
D_b(p,\delta) \lesssim \frac{p^{\delta} b^{\delta}}{\Gamma(1+\delta)}, \quad b \to \infty,
\label{eq:db_asympt}
\end{equation}
\end{lemma}
\begin{IEEEproof}
See Appendix~\ref{app:db_asym}.
\end{IEEEproof}
\begin{figure}
\centering
\includegraphics[scale=0.58]{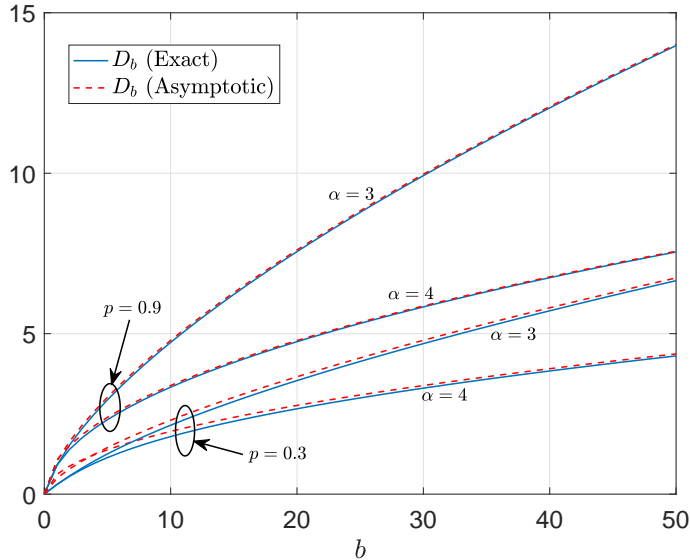}\vspace*{-1mm}
\caption{The solid lines represent the exact $D_b(p,\delta)$ as in \eqref{eq:Db}, while the dashed lines represent the asymptotic form of $D_b(p,\delta)$ as in \eqref{eq:db_asympt}.} 
\label{fig:db_asym}\vspace*{-3mm}
\end{figure}
Fig.~\ref{fig:db_asym} illustrates how quickly $D_b(p,\delta)$ approaches the asymptote.
\begin{theorem}[\textbf{de Bruijn's Tauberian theorem~{\cite[Thm.~1]{voss_2009}}}]
For a non-negative random variable $Y$, the Laplace transform $\mathbb{E}[\exp(-sY)] ~ \sim \exp(r s^{u})$ for $s \to \infty$ is equivalent to $\mathbb{P}(Y \leq \epsilon) \sim \exp(q/ \epsilon^{v})$ for $\epsilon \to 0$, when $1/u = 1/v + 1$ (for $u \in (0,1)$ and $v > 0$), and the constants $r$ and $q$ are related as $\lvert ur \rvert^{1/u} = \lvert vq \rvert^{1/v}$.
\label{th:bruijn}
\end{theorem}

\begin{theorem}[$\boldsymbol{\lambda_{\epsilon}}$ \textbf{in the high-reliability regime}]
\label{th:asym}
For $\epsilon \to 0$, the density of reliable links $\lambda_{\epsilon}$ satisfies
\begin{equation}
\lambda_{\epsilon} \sim \lambda p\exp\left(-\left(\frac{\theta p}{\epsilon}\right)^\kappa\frac{\left(\delta \lambda C' \right)^{\kappa/\delta}}{\kappa}\right), \quad \epsilon \to 0,
\label{eq:db_asym}
\end{equation}
where $\kappa = \frac{\delta}{1-\delta} = \frac{2}{\alpha -2}$ and $C' = \pi  \Gamma(1-\delta)$.
\end{theorem}
\begin{IEEEproof}
%We have 
%\begin{equation}
%\lambda_{\epsilon} =\lambda p  \mathbb{P}(P_{\rm{s}}(\theta) > 1-\epsilon).
%\label{eq:tau_eta}
%\end{equation}
%Using Thm.~\ref{th:bruijn}, we first prove that
%\begin{align}
%\mathbb{P}(P_{\rm{s}}(\theta) > 1-\epsilon) \sim \exp\left(-\left(\frac{\theta p}{\epsilon}\right)^\kappa\frac{\left(\delta \lambda C' \right)^{\frac{\kappa}{\delta}}}{\kappa}\right), \quad \epsilon \to 0,
%\label{eq:eta_asym}
%\end{align}
%which gives the desired result in \eqref{eq:db_asym}. From Lemma~\ref{lem:db_asym},
%\begin{equation}
%D_b(p,\delta) \sim p^\delta b^\delta / \Gamma(1+\delta), \quad |b| \to \infty, b\in \mathbb{C}, 
%\label{eq:asym_db}
%\end{equation}
%where $D_b(p,\delta)$ is given by~\eqref{eq:Db}. 
Let $Y= -\ln(P_{\rm{s}}(\theta))$. The Laplace transform of $Y$ is $\mathbb{E}(\exp({-sY})) = \mathbb{E}(P_{\rm{s}}(\theta)^s) = M_s(\theta)$. Using \eqref{eq:moment1} and Lemma~\ref{lem:db_asym}, we have
\begin{equation*}
M_s(\theta) \sim \exp\left(- \frac{\lambda C (\theta p)^\delta s^\delta}{ \Gamma(1+\delta)}\right), \quad |s| \to \infty.
\end{equation*}
Comparing this expression with that in Thm.~\ref{th:bruijn}, we have $r = -\frac{\lambda C (\theta p)^\delta}{ \Gamma(1+\delta)}$, $u = \delta$, $v = \delta/(1-\delta) = \kappa$, and thus $q = \frac{1}{\kappa}\left(\delta \lambda C'\right)^{\kappa/\delta}(\theta p)^{\kappa}$, where $C' = \pi  \Gamma(1-\delta)$. Using Thm.~\ref{th:bruijn}, we can now write
\begin{align}
\mathbb{P}(Y \leq \epsilon) &= \mathbb{P}(P_{\rm{s}}(\theta) \geq \exp(-\epsilon)) \nonumber \\
&\overset{(\mathrm{a})}{\sim} \mathbb{P}(P_{\rm{s}}(\theta) \geq 1-\epsilon), \quad \epsilon \to 0 \nonumber \\
 &= \exp\left(- \frac{(\theta p )^{\kappa}\left(\delta \lambda C'\right)^{\kappa/\delta}}{\kappa\epsilon^{\kappa}}\right),\label{eq:fin_be} 
\end{align}
where $(\mathrm{a})$ follows from $\exp(-\epsilon) \sim 1-\epsilon$ as $\epsilon \to 0$. Since we have 
\begin{equation}
\lambda_{\epsilon} =\lambda p  \mathbb{P}(P_{\rm{s}}(\theta) > 1-\epsilon),
\label{eq:tau_eta}
\end{equation} 
the desired result in \eqref{eq:db_asym} follows from substituting \eqref{eq:fin_be} in \eqref{eq:tau_eta}.
\end{IEEEproof}

\begin{figure}
\centering
\includegraphics[scale=0.58]{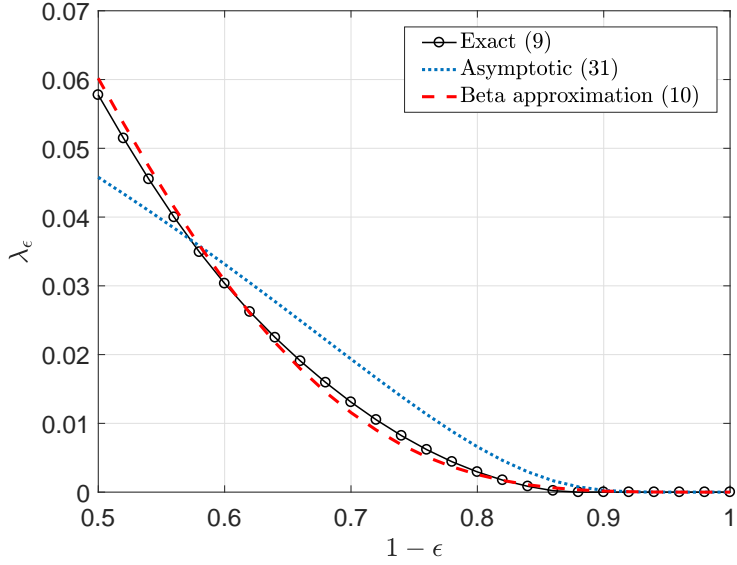}\vspace*{-1mm}
\caption{The solid line with marker `o' represents the exact expression of $\lambda_{\epsilon}$ as in \eqref{eq:exact_soc}, the dotted line represents the asymptotic expression of $\lambda_{\epsilon}$ given by \eqref{eq:db_asym} as $\epsilon \to 0$, and the dashed line represents the approximation by the beta distribution given by \eqref{eq:tau_app}. Observe that the beta approximation is quite accurate. $\theta = 0~\mathrm{dB}, \alpha = 4, \lambda = 1/2$, and $p = 1/3$.}
\label{fig:asym_eps_0}\vspace*{-3mm}
\end{figure}

For the special case of $p = 1$ (all transmitters are active), $\mathbb{P}(P_{\rm{s}}(\theta) \geq 1- \epsilon)$ in \eqref{eq:fin_be} simplifies to
\begin{align*}
\mathbb{P}(P_{\rm{s}}(\theta) \geq 1- \epsilon) \sim \exp\left(-\frac{\left(\delta \lambda C' \theta^{\delta}\right)^{\kappa/\delta}}{\kappa\epsilon^{\kappa}}\right), \quad \epsilon \to 0,
\end{align*}
in agreement with~\cite[Thm. 2]{ganti_2010} where the result for this special case was derived in a less direct way than Thm.~\ref{th:asym}. Fig.~\ref{fig:asym_eps_0} shows the behavior of \eqref{eq:db_asym} in the non-asymptotic regime and also the accuracy of the beta approximation given by~\eqref{eq:tau_app}. 
%\begin{figure}
%\centering
%\includegraphics[scale=.5]{uu}
%\caption{}
%\label{fig:3d_eps_1}
%\end{figure}

%\begin{remark}
% and are equal to $\kappa = \frac{2}{\alpha - 2}$ which is exactly the mean interference-to-signal ratio (MISR)~\cite{martin_misr} of Poisson point process for cellular networks.
%\end{remark}
We now investigate the scaling of $S(\theta,\epsilon)$ in the high-reliability regime. 
\begin{corollary}[\textbf{SOC in high-reliability regime}]
\label{cor:high_rel}
For $\epsilon \to 0$,
\begin{equation}
S(\theta,\epsilon) \sim \left(\frac{\epsilon}{\theta}\right)^{\delta}\frac{e^{-(1-\delta)}}{\pi  \delta^{\delta} \Gamma(1-\delta)},
\label{eq:cor_soc}
\end{equation}
and the SOC is achieved at $p = 1$.
\end{corollary}
\begin{IEEEproof}
For notational simplicity, let us define the rate-reliability ratio as $\rho \triangleq \epsilon/\theta$ and denote $\xi_{\rho} \triangleq \rho^{-\kappa}\frac{(\delta C')^{\kappa/\delta}}{\kappa}$ and $f_{\rho}(\lambda, p) \triangleq \lambda p \exp(-\lambda^{\kappa/\delta} p^\kappa \xi_{\rho})$. From \eqref{eq:db_asym}, we can then write $\lambda_{\epsilon} \sim f_{\rho}(\lambda, p), \quad \epsilon \to 0$, and the SOC is
\begin{equation}
S(\theta,\epsilon) \sim \sup\limits_{\lambda, p} ~f_{\rho}(\lambda,p), \quad \epsilon \to 0.
\label{eq:opt_main_soc}
\end{equation}
We can then write 
\begin{align*}
\frac{\partial f_{\rho}(\lambda,p)}{\partial \lambda} = \underbrace{p\exp\left(-\lambda^{\kappa/\delta} p^\kappa \xi_{\rho} \right)}_{> 0}\left[1 - \frac{\kappa\xi_{\rho}}{\delta}\lambda^{\kappa/\delta} p^\kappa\right].
\end{align*}
Setting $\frac{\partial f_{\rho}(\lambda,p)}{\partial \lambda} = 0$, we obtain the critical point as
\begin{align}
\lambda_0(p) = \left(\frac{\delta}{\xi_{\rho}\kappa p^\kappa}\right)^{\delta/\kappa}. 
\label{eq:opt_lam}
\end{align}
For any given $p$, the objective function is quasiconcave. Hence the optimization problem in~\eqref{eq:opt_main_soc} reduces to
\begin{align*}
S(\theta, \epsilon) &\sim\sup\limits_{p} ~f_{\rho}(\lambda_0(p),p), \quad \epsilon \to 0, \nonumber \\
&= \left(\frac{\delta}{e\kappa\xi_{\rho}}\right)^{\delta/\kappa} ~\sup\limits_{p} ~p^{1-\delta}, \quad \epsilon \to 0.
\end{align*}
Observe that $f_{\rho}(\lambda_0(p),p)$ monotonically increases with $p$ and thus attains the maximum at $p = 1$. Hence the SOC is achieved at $p = 1$ and $\lambda = \left(\frac{\delta}{\xi_{\rho}\kappa}\right)^{\delta/\kappa}$ and is given by \eqref{eq:cor_soc} after simplification.
\end{IEEEproof}
The equation~\eqref{eq:cor_soc} confirms the asymptotic bound on the SOC given in \eqref{eq:S_up_asym}.
\begin{corollary}[\textbf{The meta distribution at the SOC point}]
\label{cor:meta_high}
As $\epsilon \to 0$, the value of the meta distribution at the SOC point can be simply expressed as
\begin{equation}
\eta(\theta, \epsilon) \sim e^{-(1-\delta)}.
\label{eq:eta_high}
\end{equation}
\end{corollary}
\begin{IEEEproof}
From Cor.~\ref{cor:high_rel}, as $\epsilon \to 0$, the SOC can be expressed as
\begin{align}
S(\theta, \epsilon) \sim \lambda_{\rm{opt}}p_{\rm{opt}}\eta(\theta, \epsilon),
\label{eq:soc_eta}
\end{align}
where $\lambda_{\rm{opt}} = \lambda_0$ (given by \eqref{eq:opt_lam}) and $p_{\rm{opt}} = 1$ correspond to the SOC point as $\epsilon \to 0$. Then, comparing \eqref{eq:soc_eta} with \eqref{eq:cor_soc}, we get the desired expression of $\eta(\theta, \epsilon)$ as in \eqref{eq:eta_high}.
\end{IEEEproof}
\begin{corollary}[\textbf{The mean success probability at the SOC point}] 
As $\epsilon \to 0$, the mean success probability at the SOC point can be expressed as
\begin{equation}
p_{\rm{s,opt}} \sim 1 - \left(\frac{\epsilon}{\delta}\right)^{\delta}\Gamma(1+\delta).
\label{eq:ps_opt}
\end{equation} 
\end{corollary}
\begin{IEEEproof}
Substituting $\lambda = \lambda_0$ (given by \eqref{eq:opt_lam}) and $p = 1$ in \eqref{eq:mp_s} and using $e^{-x} \sim 1-x$ as $x \to 0$ yield the desired expression. 
\end{IEEEproof}
We now provide few remarks pertaining to the high-reliability regime.\\
\textbf{Remarks}:
\begin{itemize}
\item Letting $C_{\delta} = \left(\frac{1}{\delta}\right)^{\delta}\frac{e^{-(1-\delta)}}{\Gamma(1-\delta)}$, the density of transmitters $\lambda^{*} \triangleq S(\theta, \epsilon)$ that maximizes the density of active links that achieve a reliability at least $1-\epsilon$ behaves as 
\begin{equation}
\lambda^{*}\pi  \sim C_{\delta}\left(\frac{\epsilon}{\theta}\right)^{\delta}, \quad \epsilon \to 0.
\label{eq:mean_suc_rec}
\end{equation}
The coefficient $C_{\delta}$ depends only on $\delta$. In the practically relevant regime $1/2 \leq \delta < 1$, \textit{i.e.}, $2 < \alpha  \leq 4$, $C_{\delta} \approx 1-\delta$. In \eqref{eq:mean_suc_rec}, the left side $\lambda^{*}\pi $ is the mean number of reliable receivers in a disk of unit radius in the network. Equation \eqref{eq:mean_suc_rec} reveals an interesting trade-off between the spectral efficiency (captured by $\theta$) and the reliability (captured by $\epsilon$), where only their ratio matters. For example, at low rates, $\ln(1 + \theta) \sim \theta$; thus, a $10\times$ higher reliability can be
achieved by lowering the rate by a factor of $10$.
\item  The (potential) transmitter density $\lambda_{\rm{opt}}$ that achieves the SOC is 
\begin{equation}
\lambda_{\rm{opt}} \sim \left(\frac{\epsilon}{\theta}\right)^{\delta}\frac{1}{\pi  \delta^{\delta}\Gamma(1-\delta)},\quad \epsilon \to 0.
\label{eq:opt_lam1}
\end{equation}
Here $\lambda_{\rm{opt}}\pi$ is the mean number of (potential) transmitters in a disk of unit radius in the network that achieves the SOC. 
\item From \eqref{eq:eta_high}, it is apparent that, at the SOC point, the fraction of links that satisfy the outage constraint depends only on the path loss exponent $\alpha$, as $\delta \triangleq 2/\alpha$.
\item The mean success probability $p_{\rm{s,opt}}$ at the SOC point (given by \eqref{eq:ps_opt}) allows us to relate the SOC and the TC. Substituting $q^* = 1 - p_{\rm{s,opt}}$ in \cite[(4.29)]{weber_now}, we can express the TC as
\begin{equation*}
c(\theta,\epsilon) \sim \left(\frac{\epsilon}{\theta}\right)^{\delta}\frac{1}{\pi \delta^{\delta}  \Gamma(1-\delta)}, \quad \epsilon \to 0,
\end{equation*} 
which is the same as the expression of the optimum $\lambda$ that achieves the SOC (given by \eqref{eq:opt_lam1}). Hence $S(\theta, \epsilon) = c(\theta,\epsilon)e^{-(1-\delta)}$ if the TC framework used $p_{\rm{s}}(\theta) = p_{\rm{s,opt}}$ instead of $p_{\rm{s}}(\theta) = 1-\epsilon$ (given that $p = 1$ is optimum).
\item From Cor.~\ref{cor:high_rel}, observe that the exponents of $\theta$ and $\epsilon$ are the same. The SOC scales in $\epsilon$ similar to the TC defined in \cite{ganti_2010}, \textit{i.e.}, as $\Theta(\epsilon^{\delta})$, while the original TC defined in~\cite{weber_2005} scales linearly in $\epsilon$.
\item For $\alpha = 4$, the expression of SOC in \eqref{eq:cor_soc} simplifies to 
\begin{equation*}
S(\theta,\epsilon) ~\sim 0.154\left(\frac{\epsilon}{\theta}\right)^{1/2},  \quad \epsilon \to 0,
\end{equation*}
and the meta distribution gives $\eta \approx 0.6$. In other words, approximately $60\%$ of active links satisfy the outage constraint if $\alpha = 4$. Also, for $\alpha = 4$, the mean success probability at the SOC point is simply given by $p_{\rm{s,opt}} \sim 1 - 1.2533\sqrt{\epsilon}$ as $\epsilon \to 0$. Fig.~\ref{fig:3d_eps_0} plots $\lambda_{\epsilon}$ versus $\lambda$ and $p$ for $\epsilon = 0.007$ and $\alpha = 4$. In this case, the SOC is achieved at $p = 1$.
\begin{figure}
\centering
\includegraphics[scale=0.58]{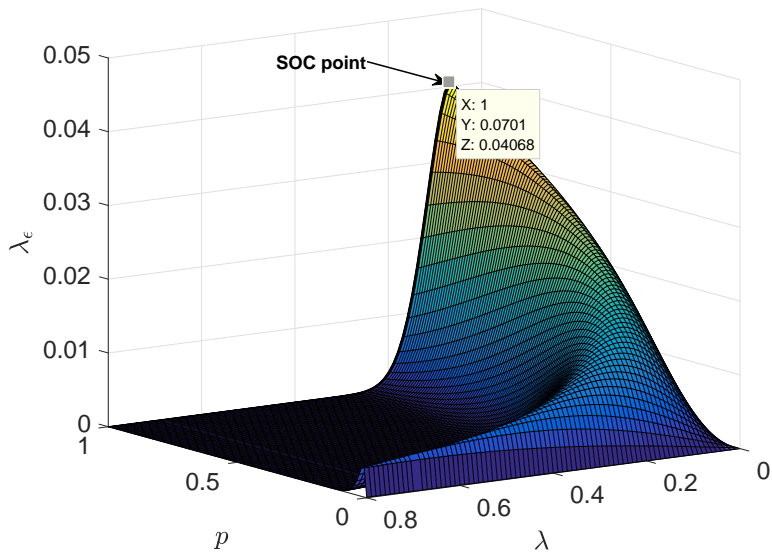}\vspace*{-1mm}
\caption{Three-dimensional plot of $\lambda_{\epsilon}$ for $\epsilon =0.007, \theta = -10~\mathrm{dB}$, and $\alpha = 4$. Observe that $p = 1$ achieves the SOC. The mean success probability $p_{\mathrm{s}}(\theta)$ at the SOC point is $0.8964$.}
\label{fig:3d_eps_0}\vspace*{-3mm}
\end{figure}
\end{itemize}

%\begin{remark}
%From the upper bound on the SOC and the exact expression of the SOC in the high-reliability regime, we can observe that the SOC scales in $R$ as $\Theta(R^{-2})$, the same as that of the TC metrics defined in \cite{weber_2005} and \cite{ganti_2010}. 
%\end{remark}

%For $\alpha = 3$ and $\alpha = 4$, \eqref{eq:db_asym} reduces to simple expressions as follow.
%\begin{corollary}[SOC for $\alpha = 3$ and $\alpha = 4$ as $\epsilon \to 0$]
%For $\alpha = 3$,
%
%$\alpha = 4$,
%\begin{equation}
%\lambda_{\epsilon} \sim \lambda p \exp\left(-\frac{\pi^{3}R^4 \theta \lambda^2 p}{4\epsilon}\right), \quad \epsilon \to 0.
%\end{equation}
%\end{corollary}
%\begin{proof}
%For $\alpha = 4$, $\delta = 1/2$. Substituting this value of $\delta$ in \eqref{eq:db_asym}, we get the desired result. 
%\end{proof}

\section{Poisson Bipolar Networks with Random Link Distances}
\label{sec:rand_link}
We now consider the case of random link distance, where the link distances are i.i.d. random variables (which are constant over time). \vspace*{-1mm}
\subsection{Network Model}
Let $R_i$ denote the random link distance between a transmitter $i$ and its associated receiver in a Poisson bipolar network. We assume that $R_i$ is Rayleigh distributed with mean $1/(2\sqrt{\mu})$ as it is the distribution of the nearest-neighbor distance in a PPP of density $\mu$~\cite{martin_dist}.\footnote{Generalizations to other link distance distributions are beyond the scope of this paper. This is because some new techniques may need to be developed, and it is unclear what other distribution to assume.} This scenario can be interpreted as the one where an active transmitter tries to communicate to its nearest receiver in a network where the potential transmitters form a PPP $\Phi_{\rm{t}}$ of density $\lambda$ and the receivers form an another PPP (independent of $\Phi_{\rm{t}}$) of density $\mu$. Similar to the deterministic link distance case, we add a receiver at the origin $o$ to the receiver PPP and an always active transmitter at location ($R_o,0$), where $R_o$ is the Rayleigh distributed link distance. Under the expectation over the point process, this link is the typical link.\vspace*{-1mm}

\subsection{Exact Formulation of the SOC}

\begin{lemma}[\textbf{$\boldsymbol{b}$th moment of the link success probability}]
\label{lem:mb_rand}
For Rayleigh distributed link distances with mean $1/(2\sqrt{\mu})$, the $b$th moment of $P_{\rm{s}}(\theta)$ is
\begin{equation}
M_b(\theta) = \frac{\mu}{\mu + \lambda \theta^{\delta}\Gamma(1+\delta)\Gamma(1-\delta)D_b(p,\delta)}.
\label{eq:Mb1}
\end{equation}
\end{lemma}
\begin{IEEEproof}
See Appendix~\ref{app:ray_dist}.
\end{IEEEproof}
For $b = 1$, using $D_1({p,\delta}) = p$, we get the expression of the mean success probability $p_{\rm{s}}(\theta)$. Moreover, for $b \in \mathbb{N}$, \eqref{eq:Mb1} represents the joint success probability of $b$ transmissions with random link distance, as obtained in \cite[(23)]{martin_diversity}.

As in the deterministic link distance case, using the Gil-Pelaez theorem, we can calculate the density of reliable links from \eqref{eq:exact_soc}, and the SOC is obtained by taking the supremum of $\lambda_{\epsilon}$ over $\lambda$ and $p$. Like the deterministic link distance case, the beta approximation is quite accurate. 

In the rest of the paper, we assume $\mu = 1$ without loss of generality.\vspace*{-1mm}
\subsection{Bounds on the SOC}
\begin{figure}
\centering
\includegraphics[scale=0.58]{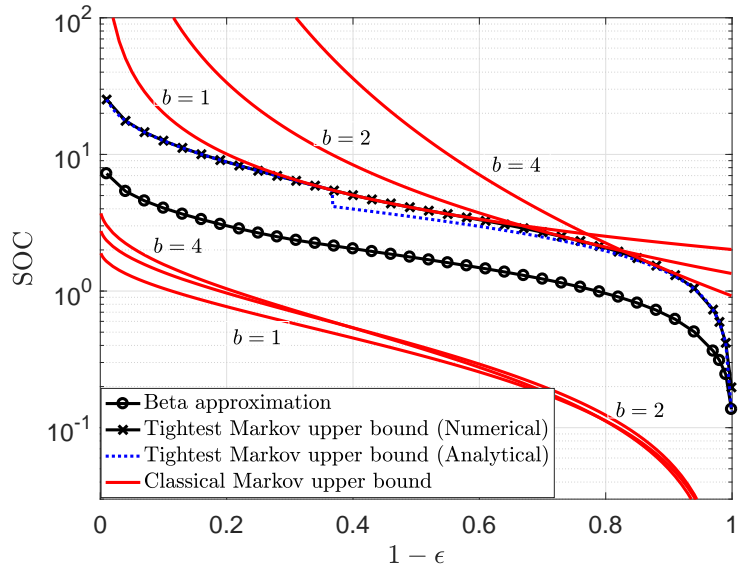}\vspace*{-1mm}
\caption{Analytical and numerical results for the SOC. The tightest Markov upper bound on the SOC obtained numerically uses \eqref{eq:soc_up_r}, which is optimized over $b$. The tightest Markov upper bound obtained analytically uses \eqref{eq:soc_up_fin_ex_r} when $\epsilon \in [0.6321, 1)$ and \eqref{eq:soc_up_r3} when $\epsilon \in (0, 0.6321)$. Observe that the analytical approximation of the SOC upper bound provides a tight upper bound for the complete range of reliability threshold $1-\epsilon$. The curve corresponding to the tightest Markov upper bound obtained analytically deviates from that obtained numerically at $1-\epsilon = 0.3679$ due to the approximation as in \eqref{eq:approx_gamma}. The classical Markov bounds are plotted using \eqref{eq:soc_up_r} for $b =1$, $b = 2$, and $b = 4$. $\theta = -10$ dB and $\alpha = 4$.}
\label{fig:upper_ray_link1}\vspace*{-3mm}
\end{figure}

\begin{theorem}[\textbf{Upper bound on the SOC}]
\label{th:soc_up_r}
For any $b > 0$, the SOC for Rayleigh distributed link distances is upper bounded as
\begin{equation}
\label{eq:soc_up_r}
 S(\theta,\epsilon) \leq \left\{
  \begin{array}{l l}
    \frac{1}{\theta^{\delta} \Gamma(1-\delta)\Gamma(1+\delta)} \frac{1}{b(1-\epsilon)^{b}}, & 0 < b \leq 1,\\
    \frac{1}{\theta^{\delta} \Gamma(1-\delta)} \frac{\Gamma(b)}{\Gamma(b+\delta)(1-\epsilon)^{b}}, & b >1.    
  \end{array} \right.
\end{equation}
%\begin{equation}
%S(\theta,x) \leq \left(\frac{-\ln x}{\theta}\right)^{\delta}\frac{e^{\delta}}{\delta^{\delta} \Gamma(1-\delta)}.
%
%\end{equation}
\end{theorem}
\begin{IEEEproof}
Again using Markov's inequality, $\eta(\theta,\epsilon)$ can be upper bounded as
\begin{align*}
\eta(\theta,\epsilon) \leq \frac{M_b(\theta)}{(1-\epsilon)^b}, \quad b > 0,
\end{align*}
where $M_b(\theta) = \frac{1}{1+ \lambda \theta^{\delta}\Gamma(1+\delta)\Gamma(1-\delta)D_b(p,\delta)}$. Hence for any $b > 0$, we have
\begin{equation*}
S \leq S_{\rm{u}},
\end{equation*}
where 
\begin{equation}
S_{\rm{u}} = \frac{1}{(1-\epsilon)^{b}} ~\sup_{\lambda,p}~ \underbrace{\frac{\lambda p}{1 + \lambda \theta^{\delta}\Gamma(1+\delta)\Gamma(1-\delta)D_b(p,\delta)}}_{A_{\lambda,p}}.
\label{eq:int_ray_1}
\end{equation}
$A_{\lambda,p}$ is maximized at $\lambda =\infty$, and it follows that
\begin{equation}
S_{\rm{u}} =  \frac{1}{\theta^{\delta}\Gamma(1+\delta)\Gamma(1-\delta)b(1-\epsilon)^{b}} ~\sup_{p}~ \frac{1}{  \:_2F_1(1-b, 1-\delta; 2; p)},
\label{eq:int_ray_11}
\end{equation}
where we have used $D_b(p,\delta) = pb\:_2F_1(1-b, 1-\delta; 2; p)$ as in \eqref{eq:db_hyper}.

Notice that the optimization problem in \eqref{eq:int_ray_11} is similar to that in \eqref{eq:sub33}. Thus, following the steps after \eqref{eq:sub33} in the proof of Thm.~\ref{thm:soc_up}, we get \eqref{eq:soc_up_r}.
\end{IEEEproof}
Similar to the deterministic link distance case (as discussed in Sec. \ref{sec:det_up} after Thm.~\ref{thm:soc_up}), from \eqref{eq:soc_up_r}, for $\epsilon \in [0.6321, 1)$, we can obtain the exact closed-form expression of the tightest Markov bound as
\begin{align}
S_{\rm{u}}^{\rm{t}} = \frac{-e\ln(1-\epsilon)}{\theta^{\delta} \Gamma(1-\delta)\Gamma(1+\delta)}.
\label{eq:soc_up_fin_ex_r}
\end{align}
For $\epsilon \in (0, 0.6321)$, we can obtain the exact tightest Markov bound numerically. Alternatively, using the approximation in \eqref{eq:approx_gamma}, we get a closed-form expression of the approximate tightest Markov bound as
\begin{equation}
\label{eq:soc_up_r3}
S_{\rm{u}}^{\rm{t}} \approx \left(\frac{-\ln(1-\epsilon)}{\theta}\right)^{\delta}\frac{e^{\delta}}{\delta^{\delta} \Gamma(1-\delta)}.
\end{equation}
As Fig.~\ref{fig:upper_ray_link1} shows, the tightest Markov upper bound on the SOC obtained analytically using \eqref{eq:soc_up_r3} deviates slightly from that obtained numerically at $\epsilon = 0.6321$ due to the approximation in \eqref{eq:approx_gamma}. As $\epsilon$ becomes smaller, $\textit{i.e.},$ $1-\epsilon$ becomes closer to $1$, the approximation \eqref{eq:soc_up_r3} becomes better. For $\epsilon < 0.2$, the gap between the approximation of the upper bound and the beta approximation is less than $0.15$ dB.

\begin{theorem}[\textbf{Lower bound on the SOC}]
The SOC is lower bounded as
\begin{align}
S(\theta, \epsilon) > \frac{(1-\sqrt{1-\epsilon})^{2}}{\epsilon \theta^{\delta} \Gamma(1+\delta)\Gamma(1-\delta)}.
\end{align}
\end{theorem}
\begin{IEEEproof}
The proof follows the proof of Thm.~\ref{th:low_det} with $M_1(\theta) = \frac{1}{1+ \lambda p \theta^{\delta}\Gamma(1+\delta)\Gamma(1-\delta)}$.
\end{IEEEproof}
Fig.~\ref{fig:upper_ray_link1} shows lower bounds on the SOC. Similar to the deterministic link distance case, the Markov lower bounds for $b \in \mathbb{R}^{+}\setminus\{1\}$ are analytically intractable.\vspace*{-1mm}

\subsection{High-reliability Regime}
\begin{theorem}[\textbf{SOC in the high-reliability regime}]
For Rayleigh distributed link distances, 
\begin{align*}
S(\theta, \epsilon) \sim \left(\frac{\epsilon}{\theta}\right)^{\delta}\frac{1}{\Gamma(1+\delta)\Gamma(1-\delta)}, \quad \epsilon \to 0,
\end{align*}
and the SOC is achieved at $p = 1$.
\end{theorem}
\begin{IEEEproof}
As in the proof of Thm.~\ref{th:asym}, let $Y = -\ln(P_{\rm{s}}(\theta))$ with its Laplace transform as $\mathcal{L}_{Y}(s) = M_{s}(\theta)$. Asymptotically,
\begin{align}
\mathcal{L}_{Y}(s) \overset{(\mathrm{a})}{\sim} \frac{A}{s^\delta}, \quad |s| \to \infty,
\label{eq:lap_taub}
\end{align} 
where $(\mathrm{a})$ follows from using $D_b(p,\delta) \sim \frac{p^\delta b^\delta}{\Gamma(1+\delta)}$ as $|b| \to \infty$ in \eqref{eq:Mb1} and thus $A = 1/(\lambda \theta^\delta p^\delta \Gamma(1-\delta))$. We claim that the expression in \eqref{eq:lap_taub} is equivalent to 
\begin{equation}
F_Y(\epsilon) \sim \frac{A\epsilon^{\delta}}{\Gamma(1 +\delta)}, \quad \epsilon \to 0.
\label{eq:lap_taub_dual}
\end{equation}
The proof that \eqref{eq:lap_taub} and \eqref{eq:lap_taub_dual} are equivalent is given in Appendix~\ref{app:lap_taub}.

As $\epsilon \to 0$, since 
\begin{align*}
F_Y(\epsilon) &= \mathbb{P}(-\ln(P_{\rm{s}}(\theta)) < \epsilon)\\ &\sim \mathbb{P}(P_{\rm{s}}(\theta) > 1-\epsilon) \\
 &= \eta(\theta, \epsilon),
 \end{align*}
the density of reliable links in the high-reliability regime can be expressed as
\begin{align}
\lambda_{\epsilon} \sim \frac{\epsilon^{\delta} p^{1-\delta}}{\theta^\delta \Gamma(1+\delta)\Gamma(1-\delta)}, \quad \epsilon \to 0.
\label{eq:tau_ran}
\end{align}
Here, $\lambda_{\epsilon}$ is independent of the density $\lambda$ of (potential) transmitters. As a result, the SOC is
\begin{align*}
S(\theta,\epsilon) \sim \frac{\epsilon^{\delta}}{\theta^\delta \Gamma(1+\delta)\Gamma(1-\delta)}~\sup_{p}~ p^{1-\delta}, \quad \epsilon \to 0.
\end{align*} 
Setting $p = 1$ achieves the SOC, which is given as
\begin{equation*}
S(\theta, \epsilon) \sim \left(\frac{\epsilon}{\theta}\right)^{\delta}\frac{1}{\Gamma(1+\delta)\Gamma(1-\delta)}, \quad \epsilon \to 0.
\end{equation*}
\end{IEEEproof}
Similar to the deterministic link distance case, only the ratio of the spectral efficiency and the reliability matters. As we observe from \eqref{eq:tau_ran}, $\lambda_{\epsilon}$ does not depend on $\lambda$. This is due to the fact that, in the high-reliability regime, the increase in $\lambda$ decreases linearly the fraction of reliable links $\eta(\theta,\epsilon)$. For example, a 2$\times$ increase in $\lambda$ decreases $\eta(\theta,\epsilon)$ by a factor of $2$. Also, the SOC is a function of just two parameters, the reliability-to-target-SIR ratio and a constant that depends only on $\delta$, \textit{i.e.}, on the path loss exponent $\alpha$. 

\section{Conclusions}    
\label{sec:conclusions}
This paper introduces a new notion of capacity, termed {\em spatial outage capacity} (SOC), which is the maximum density of concurrently active links that meet a certain constraint on the success probability. Hence the SOC provides a mathematical foundation for questions of network densification under strict reliability constraints. Since the definition of the SOC is very general, {\em i.e.}, it is not restricted to a specific point process model, link distance distribution, MAC scheme, transmitter-receiver association schemes, fading distribution, power control scheme, etc., it is applicable to a wide range of wireless networks.

For Poisson bipolar networks with ALOHA and Rayleigh fading, we provide an exact analytical expression and a simple approximation for the density of reliable links $\lambda_{\epsilon}$. The SOC can be easily calculated  numerically as the supremum of $\lambda_{\epsilon}$ obtained by optimizing over the density of (potential) transmitters $\lambda$ and the transmit probability $p$. 

In the high-reliability regime where the target outage probability $\epsilon$ of a link goes to $0$, we give a closed-form expression of the SOC which reveals
\begin{itemize}
\item the trade-off between the spectral efficiency and the reliability where only their ratio matters while calculating the SOC. 
\item insights on the scaling behavior of the SOC where, for both deterministic and Rayleigh distributed link distance cases, we show that the SOC scales in $\epsilon$ as $\Theta(\epsilon^{\delta})$.
\end{itemize} 
Interestingly, $p = 1$ achieves the SOC in the high-reliability regime. This means that with ALOHA, all transmitters should be active in order to maximize the number of reliable transmissions in a unit area that succeed with a probability close to one. Hence, in the high-reliability regime, backing off is not SOC-achieving. This happens because the reduction in the density of active links with $p$ cannot be overcome by the increase in the fraction of reliable links.

For Rayleigh distributed link distances, in the high-reliability regime, we have shown that the density of reliable links does not depend on $\lambda$ as the increase in $\lambda$ is exactly offset by the fraction of reliable links. To be precise, a $t$-fold increase in $\lambda$ decreases the density of reliable links by a factor of $t$.

As a future work, it is important to generalize the results obtained for Rayleigh fading to other fading distributions. However, since the current bounds on the SOC and the high-reliability regime results exploit a structure induced by Rayleigh fading assumption, one might need to develop new techniques depending on the fading distribution considered.

\appendices

\section{Proof of Lemma~\ref{lem:db_asym}}
\label{app:db_asym}
From \eqref{eq:Db}, we have
\begin{align}
D_b(p, \delta) &= \sum_{k = 1}^{\infty} {\binom{b}{k}}{\binom{\delta-1}{k-1}}p^k \nonumber\\
& = p \underbrace{\sum_{k = 1}^{\infty} {\binom{b}{k}}{\binom{\delta-1}{k-1}}p^{k-1}}_{A_k(p)}\label{eq:three}.
\end{align}
By Taylor's theorem,
\begin{align}
A_k(p) =  \sum_{j = 0}^{\infty} \frac{A_k^{(j)}(1)}{j!}(p-1)^j,
\label{eq:one}
\end{align}
where $A_k^{(j)}(1)$ is the $j$th derivative of $A_k(p)$ at $p = 1$. Let $(k)_j \triangleq k (k-1)(k-2)\cdot \cdot \cdot(k-j+1)$ denote the falling factorial. Then $A_k^{(j)}(1)$ can be written as 
\begin{align}
A_k^{(j)}(1) &= \sum_{k = 1}^{\infty} {\binom{b}{k}}{\binom{\delta-1}{k-1}}(k-1)_j\nonumber \\
& = \frac{\Gamma(b +\delta -j)}{\Gamma(b-j)\Gamma(1+\delta)}(\delta - 1)_j \nonumber\\
& \overset{(\mathrm{a})}{\lesssim} \frac{b^{\delta}(\delta - 1)_j}{\Gamma(1+\delta)}, \label{eq:last}
\end{align}
where $({\mathrm{a}})$ follows from $\frac{\Gamma(b +\delta -j)}{\Gamma(b-j)} \lesssim b^{\delta}$ as $b \to \infty$. From \eqref{eq:one} and \eqref{eq:last},
\begin{align}
A_k(p) \lesssim  \frac{b^\delta}{\Gamma(1+\delta)} \underbrace{\sum_{j = 0}^{\infty} \frac{(\delta - 1)_j}{j!}(p-1)^j}_{p^{\delta-1}}.
\label{eq:two}
\end{align}
From \eqref{eq:three} and \eqref{eq:two}, we get the desired result.

\section{Proof of Lemma~\ref{lem:mb_rand}}
\label{app:ray_dist}
%The proof is based on the probability generating functional (PGFL) of the relative distance process introduced in~\cite[Lemma 1]{ganti_relative}.

Let us denote the random link distance of the typical link by $R$. Then the probability density function of $R$ is $f_{R}(a) = 2\pi\mu a \exp(-\pi\mu a^2)$. Let $\|z\|$ denote the distance between a receiver and a potential interferer $z \in \Phi_{\rm{t}}$. Given $\Phi_{\rm{t}}$, the conditional link success probability $P_{\rm{s}}(\theta)$ is
\begin{equation*}
P_{\rm{s}}(\theta) = \mathbb{P}\left(\frac{hR^{-\alpha}}{I} > \theta \mid \Phi_{\rm{t}}\right) = \mathbb{E}\left(\bm{1}(h > \theta R^{\alpha}I) \mid \Phi_{\rm{t}}\right),
\end{equation*}
where 
\begin{align*}
I = \sum_{z \in \Phi_{\rm{t}}\setminus\lbrace z_o\rbrace} h_z \|z\|^{-\alpha} \bm{1}(z \in \Phi_{\rm{t}}),
\end{align*}
where $z_o \in \Phi_{\rm{t}}$ denotes the desired transmitter. Conditioning on $R$ and then averaging over fading and ALOHA results in
\begin{align*}
P_{\rm{s}}(\theta) \mid R=\prod_{z\in\Phi_{\rm{t}}\setminus\lbrace z_o\rbrace} \left(\frac{p}{1+\theta\Bigl(\frac{R}{\|z\|}\Bigr)^{\alpha}}+1-p\right) .
\end{align*}
Let $f(r) = \left(\frac{p}{1+\theta r^{\alpha}}+1-p\right)^{b}$. Then the $b$th moment of $P_{\rm{s}}(\theta)$ is
\begin{align}
M_b(\theta) &= \mathbb{E}\left[ \prod_{z\in \Phi_t\setminus\lbrace z_o\rbrace} f\left(\frac{R}{\|z\|} \right)\right] \nonumber \\
&\overset{(\mathrm{a})}{=}\mathbb{E}_R \left[\exp\left(-2 \pi \lambda\int_0^\infty t \left(1-f\left (\frac{R}{t}\right)\right)\mathrm{d} t\right)\right] \nonumber\\
&\overset{(\mathrm{b})}{=}2\pi\mu\int\limits_0^\infty  a \exp\left(-2 \pi \lambda\int\limits_0^\infty t \left(1-f\left (\frac{a}{t}\right)\right)\mathrm{d} t\right)e^{-\mu \pi a^2}\mathrm{d} a \nonumber\\
&\overset{(\mathrm{c})}{=}2\pi\mu\int\limits_0^\infty a \exp\left(-2 \pi  \lambda a^2\int\limits_0^\infty y \left(1-f\left (\frac{1}{y}\right)\right)\mathrm{d} y\right)e^{-\mu \pi a^2}\mathrm{d} a \nonumber \\
&=\frac{\mu}{\mu+2\lambda\int_0^\infty y \left(1-f(1/y)\right)\mathrm{d} y} \nonumber \\
&\overset{(\mathrm{d})}{=}\frac{\mu}{\mu + 2\lambda \underbrace{\int\limits_{0}^{\infty}\left(1-\left(1-\frac{p\theta r^\alpha}{1+\theta r^\alpha}\right)^b\right)r^{-3}\mathrm{d}r}_{F_b}},
\label{eq:Mb_ran}
\end{align}
where $(\mathrm{a})$ follows from the probability generating functional of the PPP~\cite[Chapter 4]{martin_book}, $(\mathrm{b})$ follows from the de-conditioning on $R$, $(\mathrm{c})$ follows from the substitution $y = t/a$, and $(\mathrm{d})$ follows from the substitution $y = 1/r$ and plugging $f(r)$ back. With $1$ as the upper limit of the integral and $\mu = \lambda$, \eqref{eq:Mb_ran} reduces to the expression of the $b$th moment of the success probability in a Poisson cellular network as in \cite[(27)]{martin_meta_2016}.

With $r^{\alpha} = x$, the integral in \eqref{eq:Mb_ran} can be expressed as
\begin{align}
F_b &= \frac{1}{\alpha}\int\limits_{0}^{\infty} \left(1-\left(1-\frac{p\theta x}{1+\theta x}\right)^b\right)x^{-\delta-1}\mathrm{d}x \nonumber \\
& \overset{(\mathrm{e})}= \sum\limits_{k = 1}^{\infty} {\binom{b}{k}}(-1)^{k+1} \frac{(p\theta)^{k}}{\alpha} \int\limits_{0}^{\infty}\frac{x^{k-\delta-1}}{\left(1+\theta x \right)^{k}}\mathrm{d}x \nonumber \\
& \overset{(\mathrm{f})}{=} \frac{\theta^{\delta}}{2}\frac{\pi \delta}{\sin(\pi\delta)}D_b(p,\delta),
\label{eq:fb_fin}
\end{align}
where $(\mathrm{e})$ follows from the binomial expansion of $\left(1-\frac{p\theta x}{1+\theta x}\right)^b$ and Fubini's theorem, and $(\mathrm{f})$ follows from
\begin{equation*}
\int\limits_{0}^{\infty}\frac{x^{k-\delta-1}}{\left(1+\theta x \right)^{k}}\mathrm{d}x = \theta^{\delta-k}\left[\frac{\pi}{\sin(\pi\delta)}\frac{\Gamma(k-\delta)}{\Gamma(k)\Gamma(1-\delta)}\right]
\end{equation*}
and $(-1)^{k+1}{\binom{k-\delta-1}{k-1}} = {\binom{\delta-1}{k-1}}$. Finally, substituting \eqref{eq:fb_fin} in \eqref{eq:Mb_ran} and using $\frac{\pi \delta}{\sin(\pi\delta)} \equiv \Gamma(1+\delta)\Gamma(1-\delta)$, \eqref{eq:Mb1} follows.
%The expression in \eqref{eq:rdp} is similar to the one obtained for the relative distance process for cellular networks in~\cite[Lemma 1]{ganti_relative}, except the upper limit of the integral is $\infty$ in our case instead of $1$ and $\mu = \lambda$ in \cite{ganti_relative}. 

\section{Proof that \eqref{eq:lap_taub} and \eqref{eq:lap_taub_dual} are equivalent}
\label{app:lap_taub}
The proof uses the Weierstrass approximation theorem that any continuous function $f : [t_1, t_2] \to \mathbb{R}$ can be approximated by a sequence of polynomials from above and below. 

In our case, $t_1 = 0$ and $t_2 = 1$. Thus, for any given $t > 0$, if $f(y)$ is a continuous real-valued function on $[0,1]$, for $n \geq 1$, there exists a sequence of polynomials $P_n(y)$ and $Q_n(y)$ such that 
\begin{equation}
P_n(y) \leq f(y) \leq Q_n(y) \quad \forall y\in[0,1],
\label{eq:con1}
\end{equation}
\begin{equation}
\int_{0}^{1}(Q_n(y) -f(y))\mathrm{d}y \leq t,
\label{eq:con2}
\end{equation}
and 
\begin{equation}
\int_{0}^{1}(f(y) - P_n(y))\mathrm{d}y \leq t.
\label{eq:con3}
\end{equation}
Even if $f(y)$ has a discontinuity of the first kind, we can still construct polynomials $P_n(y)$ and $Q_n(y)$ that satisfy \eqref{eq:con1}-\eqref{eq:con3}.\footnote{See \cite[Sec. 7.53]{titchmarsh_book} for the details of the construction of such polynomials.}

To prove the desired result, we first show that 
\begin{equation}
\lim_{s \to \infty} s^{\delta}\int_{0}^{\infty}e^{-sy}f(e^{-sy}) \mathrm{d}F_Y(y) = \frac{A}{\Gamma(\delta)}\int_{0}^{\infty} y^{\delta - 1}f(e^{-y})e^{-y}\mathrm{d}y.
\label{eq:des}
\end{equation}
Let $Q_n(y) = \sum_{k = 0}^{n}a_ky^k$ with $a_k \in \mathbb{R}$ for $k = 0, 1, \dotsc, n$. We then have
\begin{align*}
\limsup_{s\to\infty} s^{\delta} \int_{0}^{\infty} e^{-sy}f(e^{-sy}) \, dF_Y(y)
&\leq \lim_{s\to\infty} s^{\delta} \int_{0}^{\infty} e^{-sy}Q_n(e^{-sy}) \, dF_Y(y) \\
& = \lim_{s \to \infty}\sum_{k=0}^{n} a_k s^{\delta} \int_{0}^{\infty} e^{-(k+1)sy} \, dF_Y(y) \\
& \overset{(\mathrm{a})}{=} A \sum_{k=0}^{n} \frac{a_k}{(k+1)^{\delta}}\\
& \overset{(\mathrm{b})}{=} \frac{A}{\Gamma(\delta)} \int_{0}^{\infty} y^{\delta-1} e^{-y} Q_n(e^{-y}) \, \mathrm{d}y \\
&\overset{(\mathrm{c})}{=} \frac{A}{\Gamma(\delta)} \int_{0}^{\infty} y^{\delta-1} e^{-y} f(e^{-y}) \, \mathrm{d}y,
\end{align*}
where $(\mathrm{a})$ follows from $\underset{s \to \infty}{\lim} s^{\delta} \int_{0}^{\infty} e^{-sy} \, dF_Y(y) = A$, $(\mathrm{b})$ follows from the definition of the gamma function as $\Gamma(\delta) \triangleq \int_0^{\infty} y^{\delta-1}e^{-y}\mathrm{d}y$, and $(\mathrm{c})$ follows from the dominated convergence theorem as $n \to \infty$.

By a similar argument for $P_n(y)$, we have
\begin{align*}
\liminf_{s\to\infty} s^{\delta} \int_{0}^{\infty} e^{-sy}f(e^{-sy}) \, dF_Y(y)
&\geq \frac{A}{\Gamma(\delta)} \int_{0}^{\infty} y^{\delta-1} e^{-y} f(e^{-y}) \, \mathrm{d}y,
\end{align*}
and \eqref{eq:des} follows.

Now let 
\begin{equation}
\label{eq:binary}
f(y) = \left\{
  \begin{array}{l l}
    \frac{1}{y}, & \quad \frac{1}{e} \leq y \leq 1 \\
    0, & \quad  0 \leq y < \frac{1}{e}.\\
  \end{array} \right.
\end{equation}
Letting $s = 1/\epsilon$ in \eqref{eq:des} and using \eqref{eq:binary}, we have
\begin{align*}
\lim_{\epsilon \to 0} \epsilon^{-\delta} F_Y(\epsilon)
&= \lim_{s \to \infty} s^{\delta} \int_{0}^{\infty} e^{-sy} f(e^{-sy}) \, dF_Y(y) \\
& \overset{(\mathrm{d})}{=} \frac{A}{\Gamma(\delta)} \int_{0}^{1} y^{\delta-1} \, \mathrm{d}y \nonumber \\
&= \frac{A}{\Gamma(1+\delta)}.
\end{align*}
where $(\mathrm{d})$ follows from \eqref{eq:des} and \eqref{eq:binary}.
%\begin{align*}
%s^{\delta} \int_{0}^{\infty} e^{-sy}Q_n(e^{-sy}) \, \mathrm{d}F_Y(y)
%&= \sum_{k=0}^{n} a_k \left(s^{\delta} \int_{0}^{\infty} e^{-(k+1)sy}\mathrm{d}F_Y(y)\right) \\
%&\xrightarrow[s\to\infty]{} A \sum_{k=0}^{n} \frac{a_k}{(k+1)^b}
% = \frac{A}{\Gamma(b)} \int_{0}^{\infty} x^{b-1} e^{-x} p(e^{-x}) \, dx.
%\end{align*}

\section*{Acknowledgment}
The authors would like to thank Ketan Rajawat and Amrit Singh Bedi for their insights on the optimization problems in the paper.

\bibliographystyle{ieeetr}
\bibliography{paper}

\begin{thebibliography}{10}

\bibitem{sanket_icc17}
S.~S. Kalamkar and M.~Haenggi, ``Spatial outage capacity of {Poisson} bipolar
  networks,'' in {\em Proc. IEEE International Conference on Communications
  (ICC'17)}, (Paris, France), May 2017.

\bibitem{martin_meta_2016}
M.~Haenggi, ``The meta distribution of the {SIR} in {Poisson} bipolar and
  cellular networks,'' {\em IEEE Transactions on Wireless Communications},
  vol.~15, pp.~2577--2589, April 2016.

\bibitem{zorzi_1995}
M.~Zorzi and S.~Pupolin, ``Optimum transmission ranges in multihop packet radio
  networks in the presence of fading,'' {\em IEEE Transactions on
  Communications}, vol.~43, pp.~2201--2205, July 1995.

\bibitem{baccelli_2006}
F.~Baccelli, B.~B{\l}aszczyszyn, and P.~M{\"u}hlethaler, ``An {ALOHA} protocol
  for multihop mobile wireless networks,'' {\em IEEE Transactions on
  Information Theory}, vol.~52, pp.~421--436, February 2006.

\bibitem{baccelli_infocom}
F.~Baccelli and B.~B{\l}aszczyszyn, ``A new phase transitions for local delays
  in {MANETs},'' in {\em Proc. IEEE International Conference on Computer
  Communications (INFOCOM'10)}, (San Diego, CA, USA), pp.~1--9, March 2010.

\bibitem{weber_2005}
S.~P. Weber, X.~Yang, J.~G. Andrews, and G.~de~Veciana, ``Transmission capacity
  of wireless ad hoc networks with outage constraints,'' {\em IEEE Transactions
  on Information Theory}, vol.~51, pp.~4091--4102, December 2005.

\bibitem{ganti_2010}
R.~K. Ganti and J.~G. Andrews, ``Correlation of link outages in low-mobility
  spatial wireless networks,'' in {\em Proc. Asilomar Conference on Signals,
  Systems, and Computers (Asilomar'10)}, (Pacific Grove, CA, USA),
  pp.~312--316, November 2010.

\bibitem{yuanjie}
Y.~Wang, M.~Haenggi, and Z.~Tan, ``The meta distribution of the {SIR} for
  cellular networks with power control,'' {\em IEEE Transactions on
  Communications}.
\newblock Accepted.

\bibitem{cui_tcom}
Q.~Cui, X.~Yu, Y.~Wang, and M.~Haenggi, ``The {SIR} meta distribution in
  {Poisson} cellular networks with base station cooperation,'' {\em IEEE
  Transactions on Communications}.
\newblock Accepted.

\bibitem{martin_d2d}
M.~Salehi, A.~Mohammadi, and M.~Haenggi, ``Analysis of {D2D} underlaid cellular
  networks: {SIR} meta distribution and mean local delay,'' {\em IEEE
  Transactions on Communications}, vol.~65, pp.~2904--2916, July 2017.

\bibitem{deng2017}
N.~Deng and M.~Haenggi, ``A fine-grained analysis of millimeter-wave
  device-to-device networks,'' {\em IEEE Transactions on Communications},
  vol.~65, pp.~4940--4954, November 2017.

\bibitem{yuanjie_letter}
Y.~Wang, Q.~Cui, M.~Haenggi, and Z.~Tan, ``On the {SIR} meta distribution for
  {Poisson} networks with interference cancellation,'' {\em IEEE Wireless
  Communications Letters}.
\newblock Accepted.

\bibitem{weber_now}
S.~P. Weber and J.~G. Andrews, ``Transmission capacity of wireless networks,''
  {\em Foundations and Trends in Networking}, vol.~5, no.~2-3, pp.~109--281,
  2012.

\bibitem{martin_book}
M.~Haenggi, {\em Stochastic Geometry for Wireless Networks}.
\newblock Cambridge, U.K.: Cambridge Univ. Press, 2012.

\bibitem{martin_diversity}
M.~Haenggi and R.~Smarandache, ``Diversity polynomials for the analysis of
  temporal correlations in wireless networks,'' {\em IEEE Transactions on
  Wireless Communications}, vol.~12, pp.~5940--5951, November 2013.

\bibitem{gp_theorem}
J.~Gil-Pelaez, ``Note on the inversion theorem,'' {\em Biometrika}, vol.~38,
  pp.~481--482, December 1951.

\bibitem{bingham_1987}
N.~H. Bingham, C.~M. Goldie, and J.~L. Teugels, {\em Regular Variation}.
\newblock Cambridge, U.K.: Cambridge Univ. Press, 1987.

\bibitem{voss_2009}
J.~Voss, ``Upper and lower bounds in exponential {Tauberian} theorems,'' {\em
  Tbilisi Mathematical Journal}, vol.~2, pp.~41--50, 2009.

\bibitem{martin_dist}
M.~Haenggi, ``On distances in uniformly random networks,'' {\em IEEE
  Transactions on Information Theory}, vol.~51, pp.~3584--3586, October 2005.

\bibitem{titchmarsh_book}
E.~C. Titchmarsh, {\em The Theory of Functions}.
\newblock London, U.K.: Oxford Univ. Press, 2nd~ed., 1939.

\end{thebibliography}
\end{document}